\documentclass[aps, reprint, longbibliography, prfluids, onecolumn, superscriptaddress, floatfix]{revtex4-2}
\usepackage{bigints}

\usepackage{amssymb,amsmath}
\usepackage{textgreek}
\usepackage{subfig}
\usepackage{bigints}
\usepackage{hyperref}
\usepackage{float}
\usepackage{adjustbox}
\usepackage[table,xcdraw]{xcolor}
\usepackage{cleveref}
\usepackage[T1]{fontenc}
\usepackage[utf8]{inputenc}

\newcommand{\divv}{\vec{\nabla}\cdot}
\newcommand{\grad}{\vec{\nabla}}
\newcommand{\dt}{\partial_t}

\newcommand{\uv}[3]{\vec u_{#1}^{\, #2\, #3}}
\newcommand{\qv}[3]{\vec q_{#1}^{\, #2\, #3}}
\newcommand{\C}[2]{ (#1\cdot \grad)#2 + (#2\cdot \grad)#1 }

\newcommand{\udvec}{\vec{u}\,^{\dagger}}
\newcommand{\qdvec}{\vec{q}\,^{\dagger}}

\newcommand{\dd}{\,^{\dagger}}
\newcommand{\deltaa}{(Re_c^{-1} - Re^{-1})}

\begin{document}

\title{Sudden expansion stability thresholds modified by lateral flows}

\author{Timothee Salamon}
 \email{timothee.salamon@hotmail.fr}
 \affiliation{Laboratory of Fluid Mechanics and Instabilities, EPFL, CH-1015 Lausanne, Switzerland}

\author{Robin Debuysschère}
 \affiliation{Transfers, Interfaces and Processes, Université Libre de Bruxelles, Brussels, Belgium}
\author{Adam Chafaï}
 \affiliation{Transfers, Interfaces and Processes, Université Libre de Bruxelles, Brussels, Belgium}
\author{Benoit Scheid}
 \affiliation{Transfers, Interfaces and Processes, Université Libre de Bruxelles, Brussels, Belgium}

\author{Francois Gallaire}
 \affiliation{Laboratory of Fluid Mechanics and Instabilities, EPFL, CH-1015 Lausanne, Switzerland}

\begin{abstract}
We study the flow in a symmetric three-dimensional confined sudden expansion with lateral inflow at Reynolds number below 300 and varying  lateral-to-central flow rate ratio, using experiments, linear stability analysis, weakly nonlinear theory, and direct numerical simulations. Three distinct flow regimes are identified. Outside an intermediate band of lateral-to-central flow rate ratio, the flow undergoes a steady symmetry-breaking bifurcation above a critical Reynolds number, deflecting the central jet toward one side wall; weakly nonlinear analysis shows this bifurcation to be supercritical, excepting a very narrow parametric range. Within the intermediate band, no such critical Reynolds number exists and direct numerical simulations confirm that residual velocity asymmetries reflect the imposed geometric imperfections rather than intrinsic amplification. Fluctuations observed experimentally in the intermediate band of lateral-to-central flow rate ratio remain unexplained and warrant further investigation.
\end{abstract}
\maketitle

\section{Introduction}\label{intro}

Laminar flow past symmetric sudden expansions is a classical flow configuration that is widely studied. First studied experimentally under the assumption of a two-dimensional planar geometry \cite{Drikakis1997,Durst1974} and numerically \cite{Fearn1990}, it was shown that, even in the presence of a symmetric channel, the incompressible and Newtonian flow, spontaneously becomes asymmetric beyond a certain critical Reynolds number $Re_c$. At sufficiently low Reynolds numbers $Re$, the symmetric flow displays two recirculation zones of equal length and strength. Past the critical Reynolds $Re_c$, the symmetric solution bifurcates supercritically to a pair of steady asymmetric solutions characterised by an unequal size of the recirculation zones on opposite sides of the expansion.
Further work investigated the three-dimensionality of the flow by taking into account the influence of the side-walls \cite{Cherdron1978}, showing an increase in $Re_c$ as the lateral confinement increases, and a decrease in $Re_c$ as the expansion ratio of outlet to inlet width increases, as well as the sensitivity of the bifurcation to inlet velocity profile \cite{debuyssch}. On the other hand, the axisymmetric sudden expansion has been shown experimentally \cite{Sreenivasan1983,latornell_observations_1986,mullin_bifurcation_2009} to suddenly lose its axisymmetry and time invariance by an oscillation of the recirculation zone at the expansion, above a critical Reynolds number in the range $[600, 1500]$. This time and space symmetry breaking was not found to be due to an instability as reported by Sanmiguel-Rojas et al. \cite{sanmiguel-rojas_finite-amplitude_2012} but is most likely due to noise amplification as investigated by Cantwell et al. \cite{Cantwell}.

Controlling $Re_c$ is a key point in industry and laboratories to either exploit or delay the onset of instabilities in such systems. Sudden expansions have been used in series by Park, Song \& Jung \cite{park2009} to sort particles in suspension, by Chang $et$ $al.$ \cite{ho-hsien_chang_design_2010} to sort blood components, but also in heat exchangers and combustion \cite{hallett_flow_1984, zohir_heat_2011}. More recently, a modification to the 3D sudden expansion was proposed by Kim et al. \cite{kim2012mass} to produce mono-disperse population of lipid-Polymer nanoparticles with minimal control, by adding two lateral inlets, that span the whole height of the channel and oppositely located at the side walls, in which they inject polylactic-co-glycolic acid and keeping the central inlet in between the two inlets where they inject PEG+lipid. 

In this study, we analyse the stability of the flow of a confined sudden expansion when two lateral flows are added (see \cref{fig: scheme syringe}) with the same inlet width $\tilde w$ as the central stream. A set of critical Reynolds numbers $Re_c$ is determined experimentally for lateral to central inlet flow rate ratios $\Gamma\in[0, 3]$ (see definition below), revealing spontaneous symmetry breaking along the width of the channel for low and high flow rate ratios $\Gamma$, leading to a steady deflection of the flow. We then attempt to rationalize these experimental results using a global linear stability analysis to gain further insight into the flow behaviour, topology changes with $\Gamma$, and the nature of this steady symmetry breaking. For low $\Gamma\in[0, 0.2]$ and high $\Gamma\in[0.9, 3]$, linear stability predictions match the experimental results with an important increase of the $Re_c$ near the intermediate range of $\Gamma$. In this intermediate region of moderate $\Gamma\in[0.2,0.9]$, experiments showcase a peculiar time-dependent behaviour that linear stability fails to predict. In an attempt to rationalize the lack of instability predicted by the linear stability analysis at moderate $\Gamma$  and the discrepancy with experiments, we perform a weakly nonlinear analysis to predict the neutral curves and the influence of geometrical defects. Then, using direct numerical simulations, we investigate this intermediate region of moderate $\Gamma$ where linear stability analysis and experiments diverge to understand the time-dependent regime observed experimentally.

\section{Experimental setup and procedure}\label{M_and_m}
\subsection{Chip design}
\begin{figure}[H]
    \centering
    \includegraphics[width=0.8\linewidth]{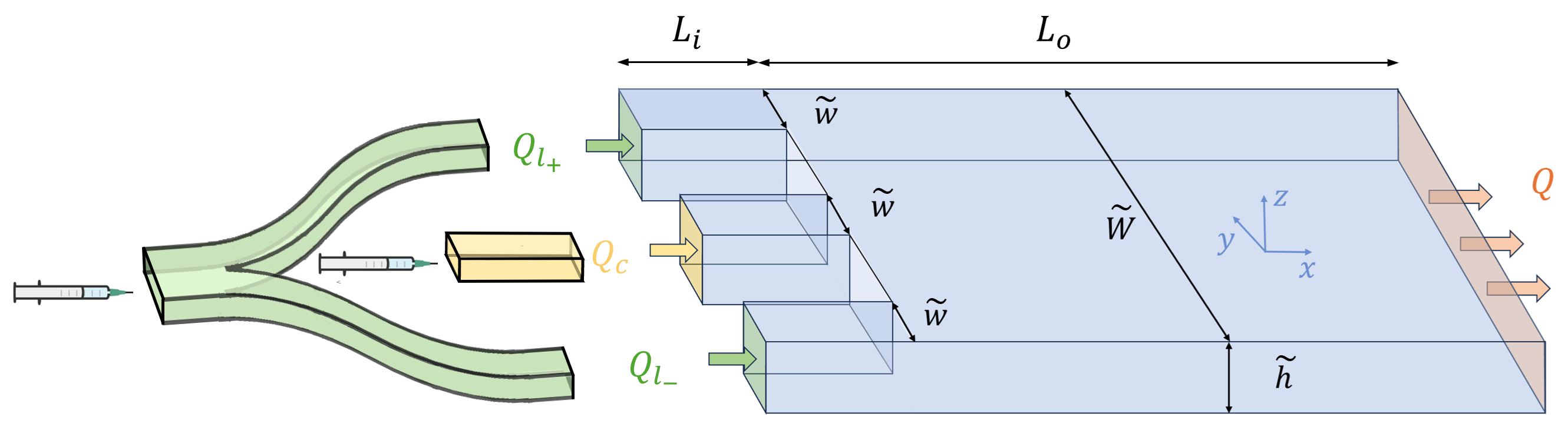}
    \caption{Schematic representation of the triple inlet system.}
    \label{fig: scheme syringe}
\end{figure}

The triple inlet channel, presented in \cref{fig: scheme syringe}, is composed of three equidistant rectangular inlets, of the same width $\tilde w$ and height $\tilde h$, distributed across a main channel of the same heigh, with a width $\tilde W$ (see fig. \ref{fig: scheme syringe} and table \ref{tab: dimensions}).  

The total flow rate $Q=Q_{l_+}+Q_{l_-}+Q_c$ defines the Reynolds number $Re = \rho \bar{U}D_h/\mu$ where $\bar U =  Q/\tilde H\tilde W$ is the mean velocity in the main chamber and $D_h=2\tilde H\tilde W/(\tilde H+\tilde W)=2/3\cdot 10^{-4}$ the hydraulic diameter, which is used as reference length scale. We use as a working fluid deionized water at room temperature $25^\circ C$ with density $\rho= 997$ kg$\cdot$m$^{-3}$ and dynamic viscosity $\mu=0.889\cdot 10^{-3}$ Pa$\cdot$s. The inlet channel length is set to 15 mm, yielding a very large length-to-confinement ratio to ensure fully developed flow before reaching the main channel. Similarly, we choose a 70 mm-long outlet channel to minimise outlet effects on the flow in the main channel as much as possible. The Reynolds number $Re$ is experimentally controlled by the total flow rate $Q=Q_l+Q_c$, while the flow rate ratio $\Gamma=Q_l/Q_c$ is controlled by imposing the flow rate at each inlet such that $Q_l = \Gamma Q/(1+\Gamma)$ and $Q_c=Q/(1+\Gamma)$. 

\begin{table}[H]
\centering
    \begin{tabular}{c|c|c|ccc|}
        \cline{2-6}
        & Symbol & Target sizes[mm] & \multicolumn{3}{c|}{Real sizes $\pm 10^{-4}$ [mm]}\\
        \hline
        \multicolumn{1}{|l|}{Width of the 3 inlets} & $\tilde w$ & 0.2 & \multicolumn{1}{l|}{$\tilde w_{l_-}=0.2030$}&\multicolumn{1}{l|}{$\tilde w_{c}=0.2020$}&\multicolumn{1}{l|}{ $\tilde w_{l_+}=0.2004$}\\
         \cline{4-6}
        \multicolumn{1}{|l|}{Depth of the channel} & $\tilde H$ & 0.4 &\multicolumn{3}{c|}{$0.4023$} \\
        \multicolumn{1}{|l|}{Width of the channel} & $\tilde W$ & 2  &\multicolumn{3}{c|}{$2.0008$} \\
        \multicolumn{1}{|l|}{Length of the inlets}& $\tilde L_i$ & 15  &\multicolumn{3}{c|}{-} \\
        \multicolumn{1}{|l|}{Length of the channel} & $\tilde L_o$ & 70  &\multicolumn{3}{c|}{-} \\
        \hline
    \end{tabular}
    \caption{Target and real sizes of the chip, where $\tilde w_{l_\pm}$ denotes the lateral inlets width and $\tilde w_c$ denotes the central inlet width. The tilde denotes dimensional quantities.}\label{tab: dimensions}
\end{table}

The chip is manufactured by engraving a 2 mm glass wafer, made of UV-grade fused silica (Siegert Wafers GmbH, Germany), using a femtosecond laser included in a FEMTOprint\textregistered (Ramat Hasharon, Israel). The focusing objective used in the FEMTOprint\textregistered was a Thorlabs LMH-20x-1064 objective (Newton, NJ, USA), and the obtained voxels have dimensions of 1.5 µm in width and 24 µm in height. After the laser exposition, the glass plate was placed in a 12 M KOH bath at 85 $^\circ$C for wet etching during 24h \cite{de2023femtosecond,amez2023instrumented}. The obtained engraved glass has a root-mean-square deviation $R_d\simeq 0.34$ $\mu$m (see appendix \ref{annex:roughness} for more details). The chip is then enclosed by a second unengraved glass wafer, held in place by bolts.

To obtain the same operating conditions and flow rates in the two lateral inlets, we designed a single inlet of $1\,\text{mm}$ diameter that contracts and then splits smoothly into two channels of the same area inside the chip (cf. fig. \ref{fig: scheme syringe}). The inlets are connected to two individually controlled low-pressure neMESYS syringe pumps (Cetoni GmbH, Germany) fitted with $50$~mL Hamilton Gastight syringes, allowing individual control of the central $Q_c$ and total lateral $Q_l=Q_{l_+}+Q_{l_-}$ flow rates. To minimise effects of the open outlet, a porous sponge is added to avoid the formation of irregular droplets.

In summary, this chip design is a carefully redesigned version of a preliminary setup described in R. Debuysschère's thesis \cite{TheseRobin} and briefly outlined in appendix \ref{appendix: prelim}. Special attention was indeed dedicated to (i) the alignment of inlet pipes to avoid secondary flow (ii) the division of the lateral channel from a single pump (iii) the length of the outlet channel (iv) alignment of the outlet with the inlet. These improvements were included to better match the numerical simulation and geometry and ensure a fair comparison. As seen in appendix \ref{appendix: prelim} the results remain, however, quite robust, in particular in parametric regions I and III (defined in section \ref{section:topology}).

\subsection{Observation and data treatment}

To visualise the flow, polystyrene $\mu$-beads (Cospheric, USA) are suspended in water with a nominal radius of $9.5$ to $11.5$ $\mu$m and density $\rho_\mu=1.070$ kg$\cdot$m$^{-3}$ such that the Reynolds number of the particles $Re_p< 1$ and the Stokes number $St_k\ll 1$. Visualisation of the flow features is performed by microscopy using a Nikon Eclipse FN1 microscope equipped with a Plan Apo 2x objective, with a Photronic Optics F5100 light source illuminating the channel from below. Images are captured with a MotionPro Y3 camera at 7300-12000 Hz for a total of more than 1000 images, 40 seconds after abruptly pushing the system to the $Re$ and $\Gamma$ of interest. Our experimental setup allows us to visualise flow only up to 20 mm downstream (30$\%$ of the channel) with a visualisation window of 12mm, which proved to be insufficient to capture the full range of flow features across the parameter range $\Gamma\in[0.2, 0.4]$.  ImageJ is then used to compute the standard deviation $\sigma(x,y)= \langle I(x,y,t)^2-\langle I(x,y,t)\rangle_t^2\rangle_t$ on the stacks of images to remove the background and recover the pathlines of the particles and to stitch images to observe the whole recirculation regions when necessary. This procedure, however, averages over the channel height and allows us to show only steady bifurcations in the channel width explicitly. This choice has been made based on previous numerical results that showed, for the range of $Re$ and $\Gamma$ considered, only steady symmetry breaking in the $y$ plane, along the width of the channel. 

 Reynolds number and flow rate ratio are then varied to find the threshold values $Re_c(\Gamma)$ for the symmetry breaking by determining the flow symmetry by sight. To better understand the flow behaviour, numerical simulations are conducted, which are described next.

 \section{Numerical approach}
\subsection{Governing equations}
\label{sub:base_state}
        We neglect the roughness of the channel by considering no-slip/no-penetration boundary conditions at the walls $\partial \Omega_w$. The flow is considered laminar and fully developed at the central $\partial \Omega_{c}$ and lateral inlets $\partial \Omega_{l_\pm}$ at an upstream distance $L_i = 2$ from the channel entrance, and the fluid is incompressible. We enforce a free stress outlet on $\partial \Omega_o$ at a downstream distance $L_{out}=40$ from the channel entrance and symmetry conditions at the symmetry planes denoted by $\partial \Omega_{s,y/z}$. Therefore, the fluid motion is governed by the incompressible Navier-Stokes equations and boundary conditions (BC):
        \begin{align}
            &\dt \uv{}{}{}  +(\uv{}{}{}\cdot\grad)\uv{}{}{} +\grad p - \frac{1}{Re} \Delta\uv{}{}{}=0, \quad  \text{in }\Omega.\nonumber\\
            &\divv \uv{}{}{} =0,\nonumber\\
            & \qquad \text{on the walls  $\partial\Omega_w$: } \vec u = 0, \label{eq:NS-full}\\
            & \qquad \text{on the outlet  $\partial\Omega_o$: } \left(p\mathbb{I}-\dfrac{\grad \vec u}{Re}\right)\cdot \vec n\cdot\vec n = 0,\nonumber\\
            & \qquad\text{on the inlets $\partial\Omega_{c/l_\pm}$: }  u(y,z) = C_{c/l_\pm}^{-1} u_{in}(y,z;y_{c/l_\pm},w_{c/l_\pm},H).\nonumber
        \end{align}
        
        Here, $u_{in}(y,z;y_{c/l_\pm},w_{c/l_\pm},H)$ is the dimensionless solution defined by eq. \ref{annex:uin} in appendix \ref{annex:inlet} for a fully developed flow in a rectangular duct of dimensionless height $H$ and width $w_{c/l_\pm}$, with $y_{c/l_\pm}$ the centers of the inlet faces. The multiplicative constants $C_{c/l_\pm}$ are defined to set the flow rate at each inlet independently, following the chosen normalisation (total flow rate of 1, lateral flow rates $hW\Gamma/2(1 + \Gamma)$, central flow rate  $Wh/(1 + \Gamma)$. The expressions of $u_{in}$ and $C_{c/l_\pm}$ are made explicit for the different computations in Appendix \ref{annex:inlet}. 

        Two different numerical approaches are used to solve these nonlinear equations with the full geometry. The first approach uses the open-source spectral element code Nek5000 (\cite{Nek5000,Patera1984}) to perform direct numerical simulations (DNS) with the channel's real dimensions (\cref{tab: dimensions}) and a flow-rate asymmetry of $ 0.5\%$ between the lateral inlets. The full three-dimensional geometry is divided into macro boxes; each macro box is then characterised by a fixed number of quadrilateral elements along the Cartesian coordinates $x$ and $y$, and is extruded with 7 layers in height, thereby forming hexahedral elements. The solution inside each of these elements is then represented in terms of $N^{th}$-order tensor-product Lagrange polynomials based on tensor product arrays of Gauss-Lobatto-Legendre quadrature points. We use the algebraic $P_N/P_{N-2}$ scheme, with $N=7$ velocity points and 5 pressure points, and the system is discretized with $30384$ spectral elements. Time integration is handled with the semi-implicit method, already implemented in Nek5000; the linear terms are treated implicitly, using a third-order backward differentiation formula, whereas the advective nonlinear term is estimated using a third-order explicit extrapolation formula. The semi-implicit scheme restricts the time step (Karniadakis et al. 1991); an adaptive time step is set to satisfy the Courant-Friedrichs-Lewy constraint, which we set to $CFL<0.1$.

        The second method uses the finite element method implemented in COMSOL software to solve the steady nonlinear problem with the symmetric target channel dimensions, while imposing an asymmetry in the lateral flow rates $\Phi$ (see \cref{eq: asymm flow rate}), using a basis of Taylor-Hood elements ($P_2$,$P_1$) for the velocity and pressure fields. The geometry is divided into macro boxes, in which we control the size of the hexahedral elements to refine the mesh locally. The steady solutions are then obtained by an iterative Newton method with a residual tolerance of $10^{-5}$.

        Finally, we also compute the linear stability properties of the nominally symmetric (in geometry and flow rate) base state as detailed next.
        
        \subsection{Base state and linear stability analysis (LSA)}
        \begin{figure}[h]
            \centering
            \includegraphics[width=.5\linewidth]{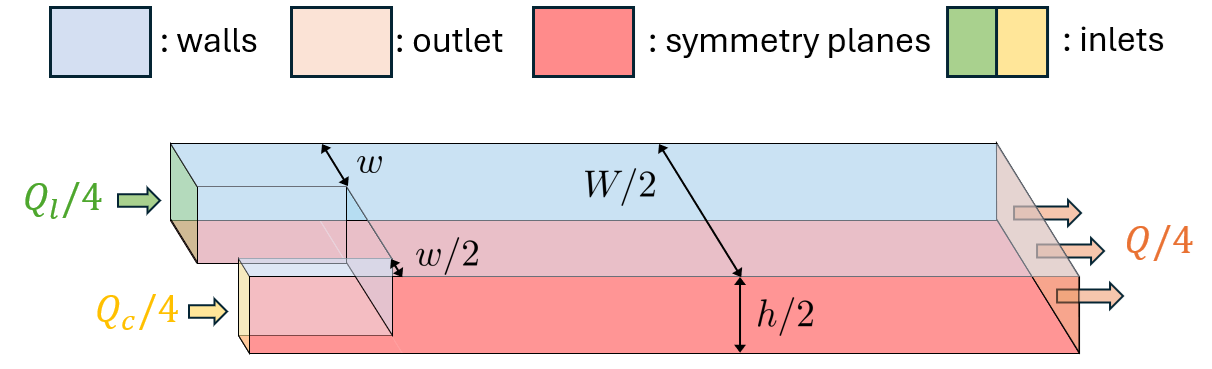}
            \caption{Scheme of the geometry used numerically by symmetry considerations.}
            \label{fig:schemes num}
        \end{figure}
        To simplify the problem and reduce the numerical cost, we analyze the dynamics of the symmetric channel using the target dimensions $\tilde w_{c,l\pm}=\tilde w=0.2$\,mm and $\tilde H= 0.4$ mm (\cref{tab: dimensions}) on only a quarter of the domain exploiting the symmetries of the system (\cref{fig:schemes num}) by using the fact that any function $f$ can be recast as a sum of a symmetric function and an antisymmetric function.  We perform a stability analysis around a steady symmetric base state $q_0(\vec x)$:
        \newline\newline\newline
        \begin{align}
            &\vec q(\vec x,t)=\begin{pmatrix} \vec u \\ p\end{pmatrix} = \vec q_0(\vec x) + \varepsilon \vec q_1(\vec x) e^{(\sigma+i\omega)t}+ \mathcal{O}(\varepsilon^2), \label{eq:modal exp}\\
            & \qquad\text{with } \vec q_0(\vec x) \in \{S_zS_y\}, \nonumber\\
            & \qquad\text{with } \vec q_1(\vec x) \in \{S_zS_y,A_zS_y,S_zA_y,A_zA_y\}. \nonumber\\
        \end{align}
        
        In the equations above, $\sigma$ is the growth rate, $\omega$ the angular frequency, and $S_i/A_i$ ($i=y,z$) denotes symmetry/antisymmetry with respect to the symmetry plane with normal $i$. Remark that if $\sigma$ is positive, then the base state is unstable along the direction $q_1$, whereas if $\sigma$ is negative, the base state is stable. 
        
        These symmetry conditions allow us to compute an ad hoc symmetric base state and then study the linear stability of each symmetry-breaking independently. To ease the notation, we use $C(\uv{k}{}{},\uv{j}{}{})=\C{\uv{k}{}{}}{\uv{j}{}{}}$ to denote the advection operator  of velocities at two different order $k$ and $j$.
        
        \begin{align}
            &\qquad\divv \uv{0}{}{} =0,\qquad \frac{1}{2}C(\uv{0}{}{},\uv{0}{}{}) +\grad p_0 - \frac{1}{Re} \Delta \uv{0}{}{}=0, \quad  \forall x\in \Omega.\\\nonumber\\
            & \qquad\qquad\uv{0}{}{}\big|_{\partial\Omega_w} = 0,\qquad  \uv{0}{}{}\big|_{\partial\Omega_{c/l_\pm}} = C^{-1}_{c/l_\pm} u_{in}(y,z;y_{c/l_\pm},w,H),\\
            & \qquad\qquad \left(p_0\mathbb{I}-\frac{\grad \uv{0}{}{}}{Re}\right)\cdot\vec n \big|_{\partial\Omega_o}=0, \qquad u_{\perp}\big|_{\partial\Omega_{s,y/z}} = \partial_\perp \vec{u}_{\parallel}\big|_{\partial\Omega_{s,y/z}}=0.
        \end{align}
        
        In the equations above, $u_\perp/u_\parallel$ and $\partial_\perp/\partial_\parallel$ denote the normal/parallel velocity(ies) and derivative(s) with respect to the plane of consideration. 
        The base state is then sought using a finite element code in COMSOL using Taylor-Hood elements with an iterative Newton method, such that the L$^2$ norm of the residual is lower than $10^{-5}$, as a steady and fully symmetric $S_yS_z$ solution to the incompressible Navier-Stokes equations.
    
        Once the steady base state is computed, we can seek symmetry-breaking perturbations that, in the linear stability framework, grow onto the base state ($\sigma>0$). Using the modal expansion defined in \cref{eq:modal exp}, one can define a generalised eigenvalue problem for $\vec q_{1}(\vec x)$ by linearizing the equations around the steady and fully symmetric base state $q_0(\vec x)$:

        \begin{align}
            & (\sigma + i\omega)\uv{1}{}{}+ C(\uv{0}{}{},\uv{1}{}{}) +\grad p_1 - \frac{1}{Re} \Delta\uv{1}{}{}=0, \quad  \forall x\in \Omega,\label{eq:lin1}\\
            & \divv \uv{1}{}{} =0,\label{eq:lin2}\\
            &\qquad\text{with BC: }\quad\uv{1}{}{}\big|_{\partial\Omega_w} = 0, \qquad \uv{0}{}{}\big|_{\partial\Omega_{c/l_\pm}} = 0,\nonumber\\
            & \qquad \qquad \qquad \qquad \qquad \qquad \left(p_1\mathbb{I}-\frac{\grad \uv{1}{}{}}{Re}\right)\cdot\vec n \big|_{\partial\Omega_o}=0,\label{eq:lin3} \\
            &\qquad \qquad\qquad\qquad\qquad\qquad  \text{and symmetry conditions at } \partial\Omega_s.\nonumber
        \end{align}
        A more compact notation for equations (\ref{eq:lin1}-\ref{eq:lin2}) is: 
        \begin{align}
            &  \left[(\sigma + i\omega)\hat B + \hat L \right]\vec q_1 = 0,\label{eq:gen eig 1st ord}\\
            &\text{with:}\qquad \hat B = \begin{pmatrix}\mathbb{I}_{\vec u} & 0\\ 0 & 0\end{pmatrix},\qquad \hat L = \begin{pmatrix} C(\uv{0}{}{}, \ldots) - \frac{\Delta}{Re} & \grad\\ \grad^T& 0 \end{pmatrix}.\nonumber
        \end{align}
        Eigenvalues and eigenstates of equation (\ref{eq:gen eig 1st ord}) are computed using the ARPACK package with a Krylov subspace of dimension 20 on the same mesh and with the same discretization as the base state. Instead of doing a single eigenvalue search with a larger number of eigenvalues, hence a larger Krylov subspace, we do multiple searches around different shifts of the eigenvalues to compute the eigenspectra. Repeating this procedure at multiple $Re$ for a given $\Gamma$ below and above the threshold one can infer a critical Reynolds number $Re_c(\Gamma)$ at which the growth rate changes sign, which is described in appendix \ref{annexe: LSA WNL}.  
        
        To further reduce the computational cost and get insights on the $\Gamma$ and $Re$ dependence of flow, we used the mean of a weakly nonlinear stability analysis (WNL) to obtain predictions of the neutral curves $\sigma(Re_c,\Gamma_c)=0$, inform us on the nature of the bifurcation (sub/super-critical) and give us a prediction of the saturated solution to the fully nonlinear Navier-Stokes equations.
        
        \subsection{Weakly nonlinear stability analysis (WNL)}
\label{sub:WNL}
The idea behind this procedure is to consider the growing symmetry-breaking perturbations obtained from linear stability analysis of the steady and symmetric base state, and to evaluate their nonlinear feedback on the base state as well as on their own evolution, to predict the perturbation growth and saturation near the critical point ($Re_c$, $\Gamma_c$) while accounting for base-state modifications induced by variations of $Re$ and $\Gamma$. In this approach, the system dynamics are reduced to the evolution of the base state under the influence of these perturbations.

We first have to define a distance measure in the parametric space ($Re$, $\Gamma$) which does not induce further non-linearities. A natural choice for the definition of this neighbourhood:

\begin{equation}
   \varepsilon^2\delta =  \dfrac{1}{Re_c}-\dfrac{1}{Re}, \qquad   \varepsilon^2 f =  \dfrac{1}{1+\Gamma_c}-\dfrac{1}{1+\Gamma}=-\left(\dfrac{\Gamma_c}{1+\Gamma_c}-\dfrac{\Gamma}{1+\Gamma}\right),\quad \varepsilon\ll1.
\end{equation}

This choice then defines a slow time scale $T=\varepsilon^2t$ which comes from the Taylor expansion of the perturbation in the neighbourhood $(Re_c,\Gamma_c)$:
\begin{equation}
    \vec q_1 e^{(\sigma+i\omega t)}\big|_{Re,\Gamma} = \vec q_1  \underbrace{\left[1+\underbrace{\varepsilon^2t}_{T} (\delta\partial_{\delta}\sigma + f\partial_f\sigma ) +\mathcal{O}(\varepsilon^4t)\right]_{Re_c,\Gamma_c}}_{A(T)} = A(T) \vec q_1 +\mathcal{O}(\varepsilon^4t)
\end{equation}

Using the Hilbert product as the norm $\langle\vec q, \vec g\rangle = \int_\Omega \vec q^{*T} \vec g\,d\omega$, we can define an adjoint operator and an adjoint mode $q^\dagger$, which are specific to this norm, such that:

\begin{align}
       0&= \langle \left((\sigma+i\omega)\hat{B} +\hat{L}\right)\vec q_1, \qdvec  \rangle =\langle \vec q_1, \left((\sigma+i\omega)\hat{B} +\hat{L}\right)^\dagger\qdvec  \rangle, \nonumber  \\
    & =\langle \vec q_1, \left((\sigma-i\omega)\hat{B} +\hat{L}^\dagger\right)\qdvec  \rangle. \label{eq:adjoint}\\
    &\Longleftrightarrow (\sigma-i\omega)\hat{B}\qdvec + \hat{L}^\dagger \qdvec =0, \qquad \hat{L}^\dagger = \begin{pmatrix}(\grad \uv{0}{}{})^T - \uv{0}{}{}\cdot\grad (\ldots) -\dfrac{\Delta}{Re} & \grad \\ 0 & \grad^T \end{pmatrix},\nonumber\\
    & \udvec\big|_{\partial\Omega_w\cup\partial\Omega_i} =0,  \quad \left(p^\dagger\mathbb{I}-\frac{\grad \udvec}{Re}\right)\cdot\vec n|_{\partial \Omega_o} = -(\uv{0}{}{}\cdot \vec n)\udvec|_{\partial \Omega_o},\nonumber\\
      &\quad \text{and same symmetry conditions as the direct mode on } \partial \Omega_s.\nonumber
\end{align}
As we will see later, the only symmetry-breaking perturbations found within the scope of this work break the $y$-symmetry in a steady manner ($\omega=0$), which greatly simplifies the equations. At second order, using the linearity of the linearized equations, we find:
\begin{align}
    &\qquad \qquad\qv{2}{}{} = A^2\qv{2}{A^2}{}+ \delta\qv{2}{\delta}{}+ f\qv{2}{f}{},\quad \qv{2}{i}{}\in{S_yS_z}, \nonumber\\
    &\hat{L} \qv{2}{A^2}{} = -C(\uv{1}{}{},\uv{1}{}{}), \qquad \text{with homogeneous BC on }\partial\Omega, \nonumber\\
    &\hat{L} \qv{2}{\delta}{} = -\Delta\uv{0}{}{}, \qquad\qquad\,\,\,\,\,\text{with homogeneous BC on }\partial\Omega,\nonumber\\
    &\hat{L} \qv{2}{f}{} = 0, \qquad\uv{2}{f}{}=\begin{cases} -hWu_{in}
    (y,z;y_{c},w,H)\quad \text{   on }\partial\Omega_{c},\\ \frac{hW}{2}u_{in}(y,z;y_{l_\pm},w,H) \quad\,\,\text{   on }\partial\Omega_{l_\pm}, \end{cases} \text{and homogeneous BC on }\partial\Omega\backslash \partial\Omega_i, \nonumber
\end{align}
which can be easily inverted due to the symmetry of $\qv{2}{i}{}$ which lifts the singularity of the operator $\hat L$ ($\hat L \qv{1}{}{}=0$ at $Re_c$). However, at third order, we recover an equation $\hat L \qv{3}{}{}= \vec F_3$ where, the forcing term, $F_3$ has the symmetries of $\qv{1}{}{}$. Due to the singularity of $\hat L$ one needs the imposition of a solvability condition (or Fredholm alternative) which is done by using the fact that $\qdvec$ is in the kernel of the adjoint operator $\hat L\dd$ (\cref{eq:adjoint}). This solvability condition enforces the evolution of $A(T)$ by a so called amplitude equation:

\begin{align}
    0=\langle\hat{L}^\dagger \udvec , \qv{3}{}{}\rangle&=\langle \udvec ,\hat{L} \qv{3}{}{}\rangle= \langle\udvec, \vec{F}_3\rangle,\nonumber\\
    \Longleftrightarrow \partial_T A \langle\qdvec, \hat B \qv{1}{}{}\rangle =& -A^3\underbrace{\langle\udvec, C(\uv{1}{}{},\uv{2}{A^2}{})\rangle}_{\mu\langle\qdvec, \hat B \qv{1}{}{}\rangle} - \delta A \underbrace{\langle \udvec, C(\uv{1}{}{},\uv{2}{\delta}{})+\Delta\uv{1}{}{} \rangle}_{\lambda\langle\qdvec, \hat B \qv{1}{}{}\rangle}- f A \underbrace{\langle \udvec, C(\uv{1}{}{},\uv{2}{f}{})\rangle}_{\alpha\langle\qdvec, \hat B \qv{1}{}{}\rangle},\nonumber\\
    &\partial_TA = -\mu A^3- (\lambda \delta + \alpha f)A,\label{eq: amp_eq_A} \\
    \underset{ a= \varepsilon A}{\Longleftrightarrow} \quad\frac{d}{dt}a = -\mu a^3 + &\left(\underbrace{-\lambda \deltaa}_{\sigma(Re,\Gamma_c)} \underbrace{-\alpha \left((1+\Gamma_c)^{-1}-(1+\Gamma)^{-1}\right)}_{\sigma(Re_c,\Gamma)}\right)a. \label{eq: amp_eq_a}
\end{align}

It is interesting to remark that for $a\ll 1$ one can neglect the nonlinear term $\mu a^3$ in equation \ref{eq: amp_eq_a} and would recover an approximation of the linear stability result: $a=e^{-(\lambda \varepsilon^2\delta +\alpha \varepsilon^2 f)t}\simeq e^{\sigma t}$. 
By definition, the neutral curves are the boundary of marginal stability ($\sigma=0$), which therefore can be approximated using that $\sigma \simeq (-\lambda \varepsilon^2\delta -\alpha \varepsilon^2 f) +\mathcal{O}(\varepsilon^4)$, their approximate parametric equation is then:

\begin{equation}
\deltaa\lambda +\left((1+\Gamma_c)^{-1}-(1+\Gamma)^{-1}\right)\alpha=0.    
\end{equation}

Keeping the nonlinear term allows us to predict the saturation amplitude $a$:
\begin{equation}
    a\underset{t\rightarrow \infty}{=}\sqrt{-\dfrac{\lambda}{\mu}\deltaa -\dfrac{\alpha}{\mu}\left((1+\Gamma_c)^{-1}-(1+\Gamma)^{-1}\right)}.
\end{equation}

Therefore, at a given $\Gamma_c$ it predicts a supercritical bifurcation if sign$(\lambda)=-$sign$(\mu)$ and a supercritical bifurcation in the case of  sign$(\lambda)=$sign$(\mu)$ in which case one would have to go to at least 5$^{\text{th}}$ order for the amplitude equation to capture the existence of 5 solutions which is detailed in appendix \ref{annexe:amp_eq}.

We can then further increase the complexity of the amplitude equation to capture asymmetries of the lateral inlets. As a model, we consider a simple asymmetry in the flow rate in the lateral inlets:

\begin{equation}
    u|_{\partial \Omega_{l_\pm}} = (1+\text{sgn}(y)\varepsilon^3\phi)C^{-1}_{l_\pm} u_{in}(y,z;y_{l_\pm},w,H). \label{eq: asymm flow rate}
\end{equation}

This adds an imperfection term $$\theta=\dfrac{\Gamma}{2(1+\Gamma)}\bigointss_{\partial\Omega_{l_+}} \left(p^{\dagger}-\dfrac{\partial_x u^\dagger}{Re}\right) C^{-1}_{l_+} u_{in}(y,z;y_{l_+},w,H)$$ to the amplitude equation \ref{eq: amp_eq_A}:

\begin{equation}
    \partial_TA = -\mu A^3- (\lambda \delta + \alpha f)A + \phi \theta.
\end{equation}

\section{Results}
\subsection{Symmetry preserving states}
\label{section:topology}
\begin{figure}[H]
    \centering
    \subfloat[$\Gamma$=0.1]{\includegraphics[width=1\linewidth]{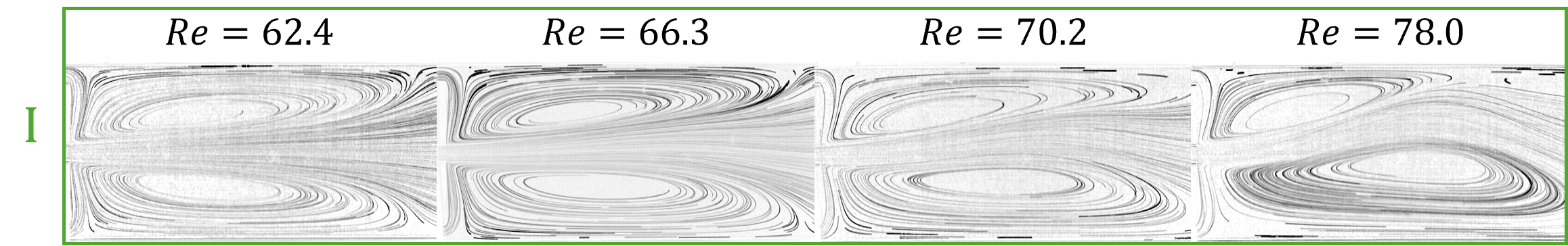}\label{fig: FRR 01}} \\
    \subfloat[$\Gamma$=0.7]{\includegraphics[width=1\linewidth]{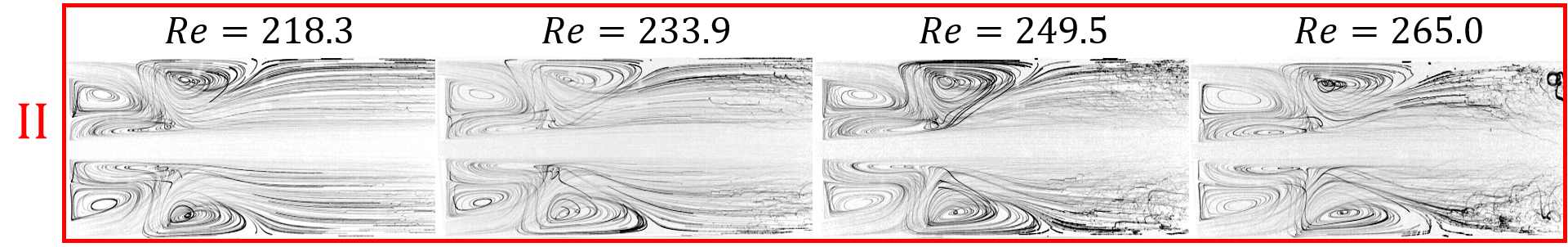}\label{fig: FRR 07}}\\
    \subfloat[$\Gamma$=1.25]{\includegraphics[width=1\linewidth]{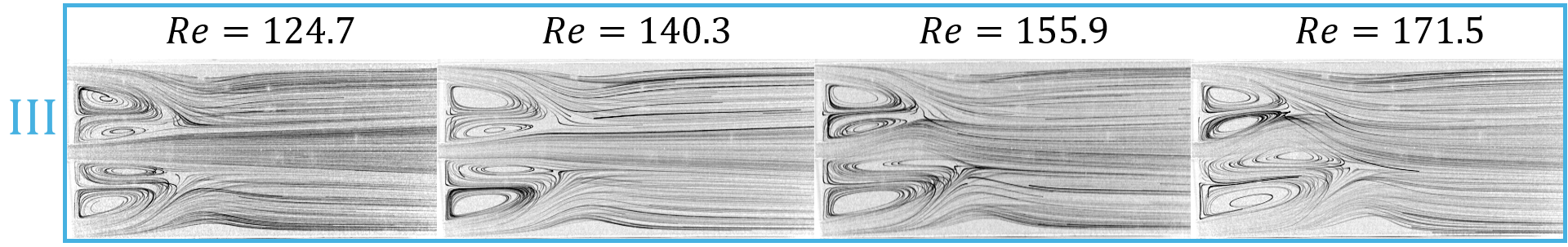}\label{fig: FRR 125}}
    \caption{Height averaged pathlines of the seeded particles in the channel at representative $\Gamma$ below and above the critical point $Re_c$. Images here are stretched in the vertical direction by an arbitrary stretching factor for the sake of clarity and reading.}
    \label{fig:bifurcation}
\end{figure}

Before diving into the results, let's recall that $\Gamma = Q_l/Q_c$, where $Q_l$ is the combined flow rate of the two lateral inlets and $Q_c$ is the flow rate of the central one. To take a few examples, this means that $\Gamma=2$ denotes equal flow in all the inlets, while at $\Gamma=1$ the central jet is twice stronger than the lateral ones, and at $\Gamma =0$ there is only the central jet. Fixing $\Gamma$ and varying $Re$ we observe that under a certain $Re_c$ the flow is symmetric and steady (\cref{fig:bifurcation}, left columns), while above the critical point the flow suddenly either bifurcates to an asymmetric steady state or starts to display a form of unsteadiness (\cref{fig:bifurcation}, right figures). 

Before describing these bifurcated states, we first present the complexity of the symmetric flow and its evolution under changes of $\Gamma$ while staying under the frontier $Re_c(\Gamma)$, the neutral curve. As shown in figure \ref{fig:bifurcation}, the flow, below the neutral curve, has distinct features depending on $\Gamma$. These flow features can be categorised into three main regions that will be made more explicit later using the numerical solutions that will give us more information on the 3D effects, presented here in order of increasing complexity:
\begin{itemize}
    \item \textbf{Region I ($0\leq \Gamma\lesssim0.2$):} In this low $\Gamma$ region $Q_c\gg Q_l$. The three jets first merge in a central jet at a distance of the order of the width of the inlets. This central jet then slowly diffuses across the entire width of the main channel, separated from the walls by \textbf{two} opposite, elongated, seemingly closed recirculation zones (see \cref{fig: FRR 01} first image).  
    \item \textbf{Region III ($1\lesssim \Gamma\leq 3$):} In this high $\Gamma$ region $Q_c\sim Q_l$. The three jets are separated by two main symmetric recirculation regions, each composed of two counter-rotating closed recirculation regions attached to the upstream wall separating the inlets, creating altogether \textbf{four} recirculations/backflow zones. The three jets then diffuse to invade the whole width of the main channel (see \cref{fig: FRR 125} first image).  
    \item \textbf{Region II ($0.2\lesssim \Gamma\lesssim1$):} In this moderate $\Gamma$ region $Q_c> Q_l$. The three jets are separated by two symmetric pairs of backflow regions, each attached to the upstream wall that separates the inlets. The three jets then merge, creating two other symmetric backflow regions attached to the side walls of the main channel creating altogether \textbf{four} to \textbf{six} recirculation/backflow region. This second pair of backflow regions appears to possess a peculiar flow topology, where some particles seem to travel back to the first set of backflow regions (see \cref{fig: FRR 07} first image).
\end{itemize}

\begin{figure}[H]
    \centering
    \includegraphics[width=\linewidth]{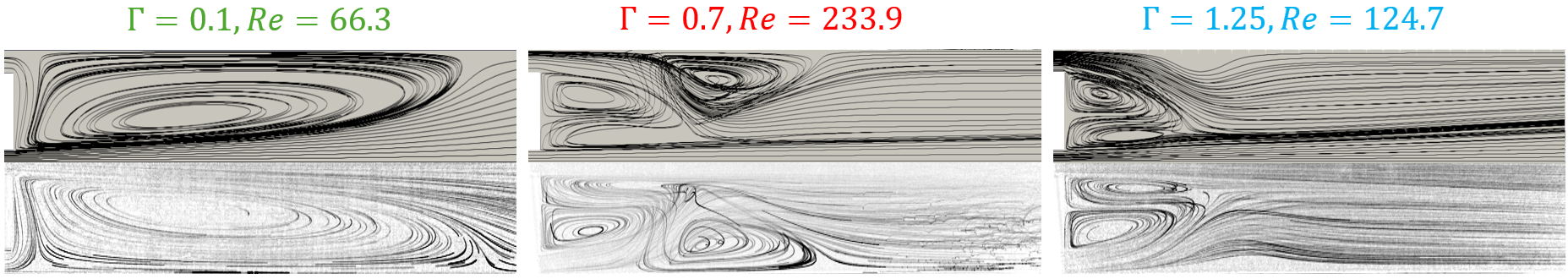}
    \caption{Qualitative comparison between experimental height-projected pathlines (bottom half of the pictures) and numerical base state $\vec u_0$ height-projected streamlines at three representative $\Gamma$ near the experimental approximate bifurcation.}
    \label{fig:comp exp LSA}
\end{figure}

Using the symmetric base flows computed as part of the LSA, we can qualitatively compare our experimental symmetry-preserving flows to the steady symmetric base states $\vec u_0$ obtained and gain more insight into the 3D nature of the flow. Figure \ref{fig:comp exp LSA} shows good qualitative agreement between the height projected streamlines computed and the pathlines experimentally observed on the mid plane of the channel of the experimental and numerical data. 

\begin{figure}
    \centering
    \subfloat[$\Gamma = 0.1, Re = 67$]{\includegraphics[width=.5\linewidth]{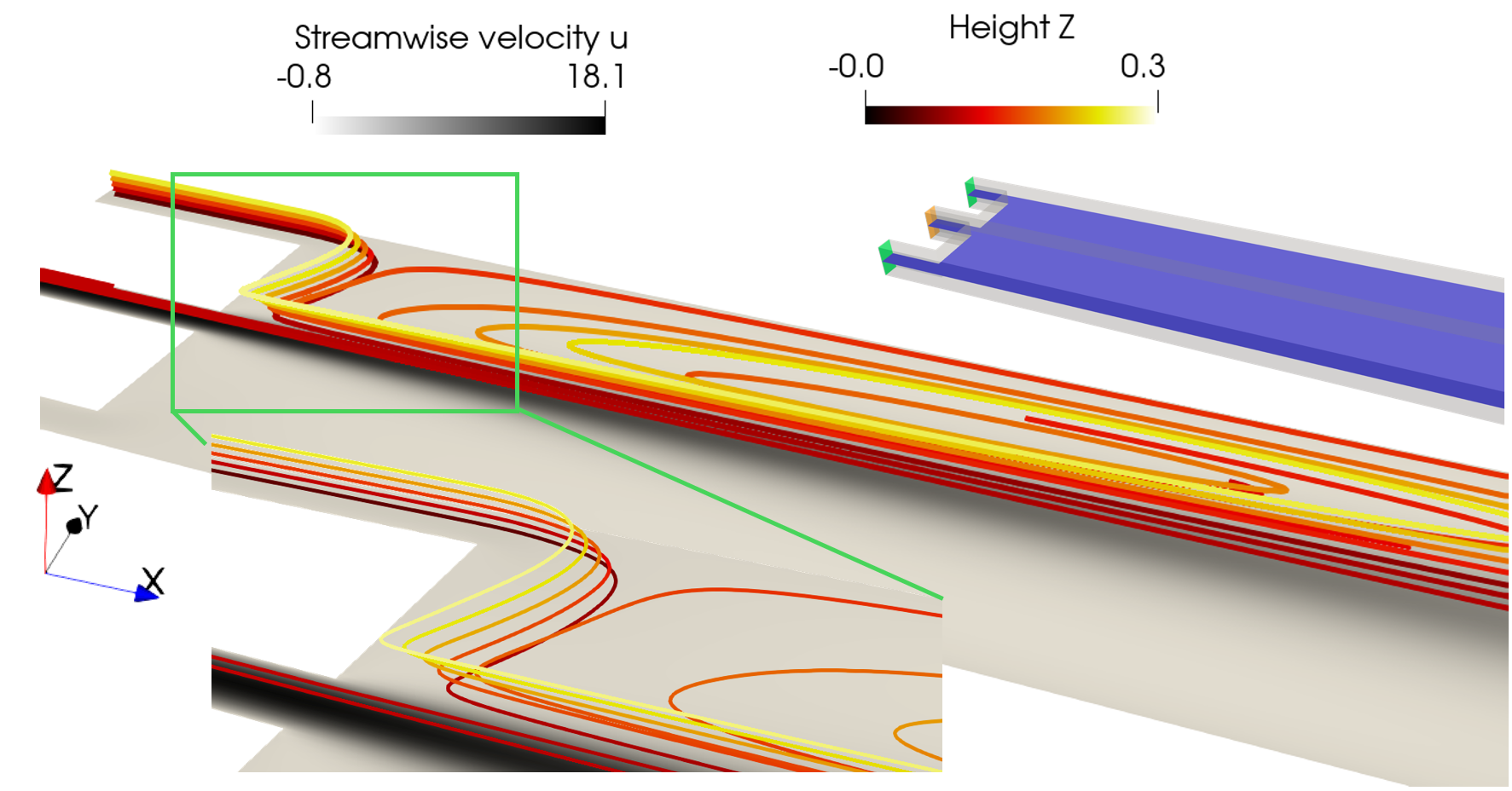}\label{fig:stream 0.1}}
    \subfloat[$\Gamma = 0.2, Re = 141$]{\includegraphics[width=.5\linewidth]{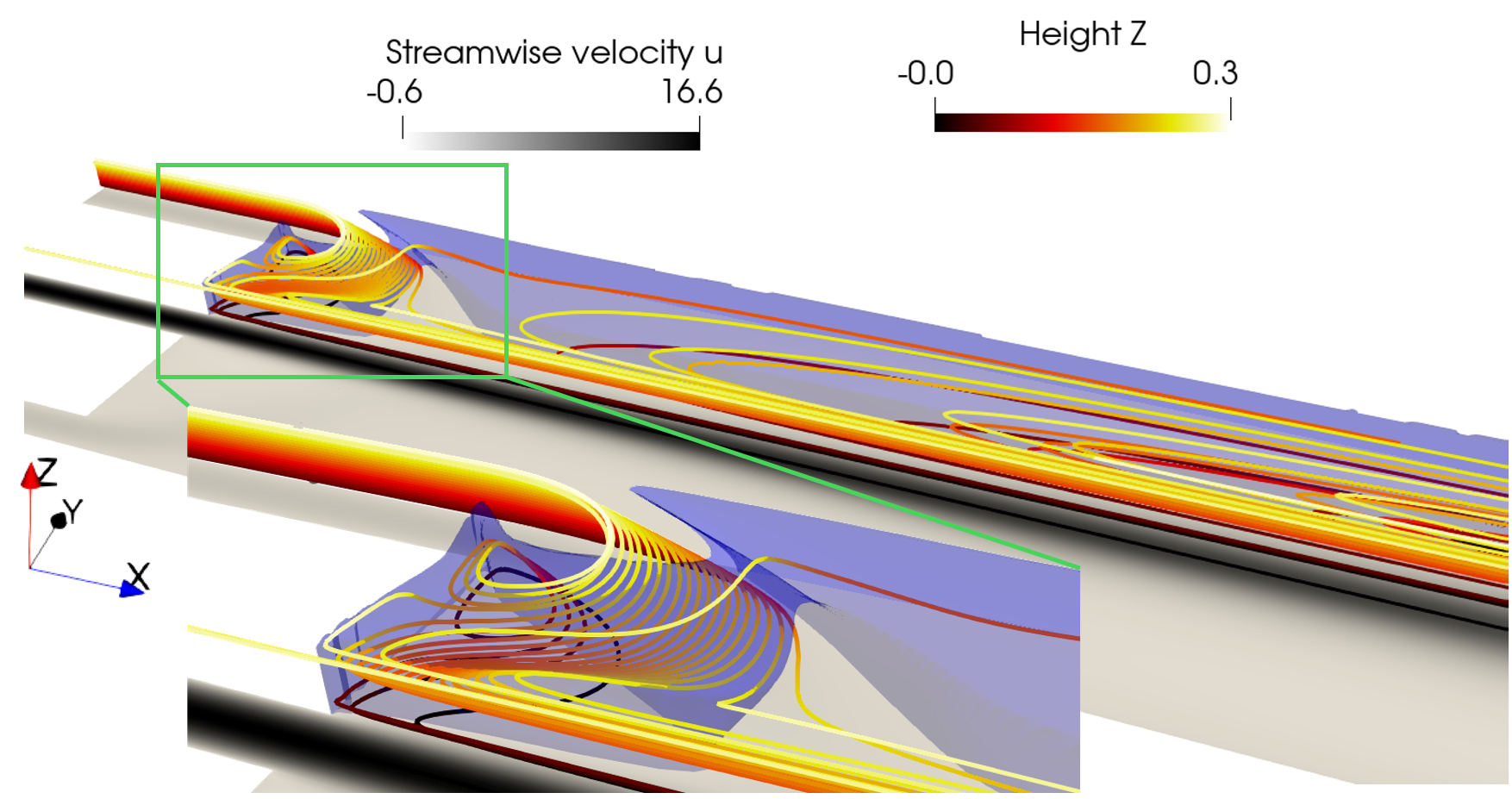}\label{fig:stream 0.2}}\\
    \subfloat[$\Gamma = 0.4, Re = 188$]{\includegraphics[width=.5\linewidth]{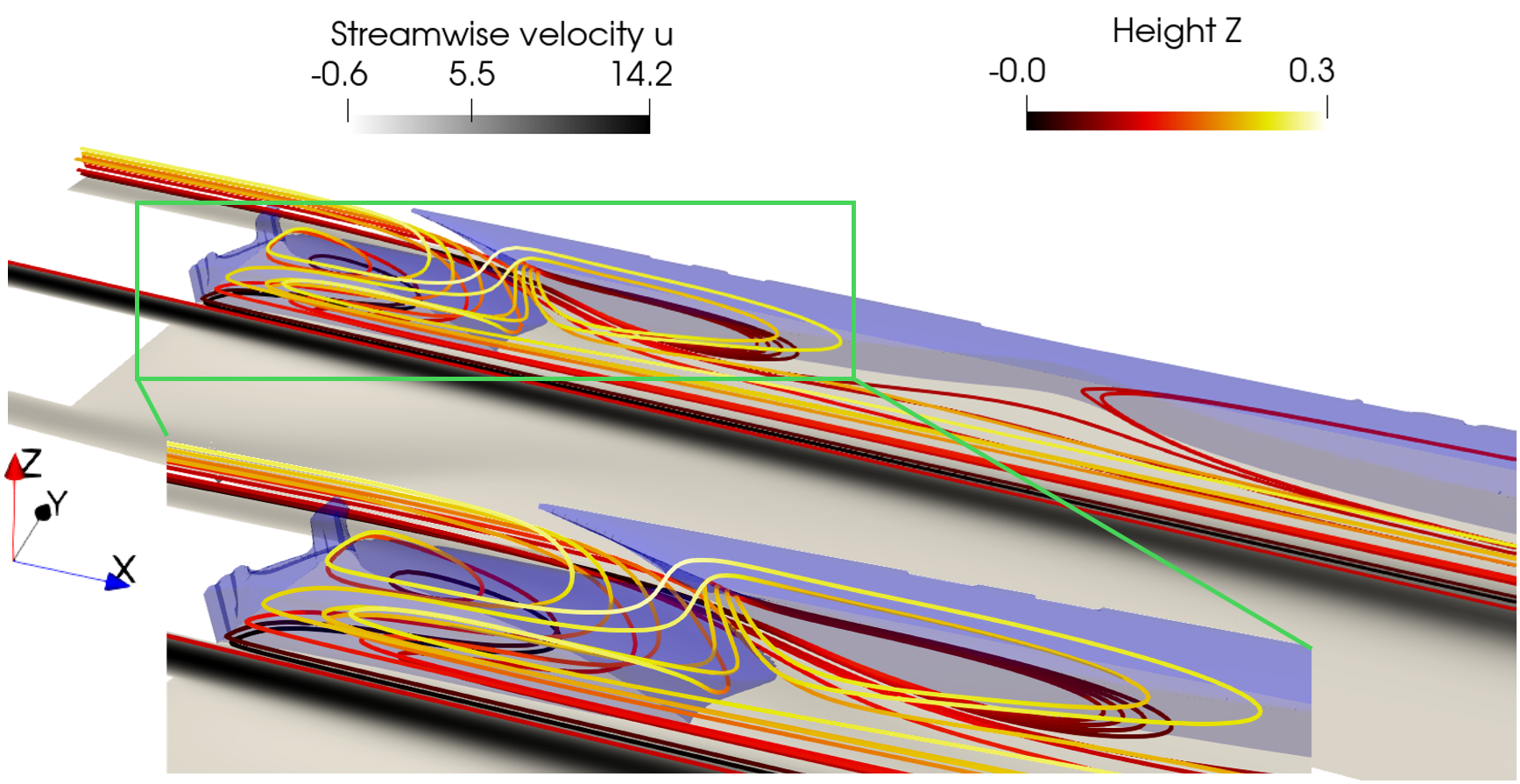}\label{fig:stream 0.4}}
    \subfloat[$\Gamma = 0.7, Re = 235$]{\includegraphics[width=.5\linewidth]{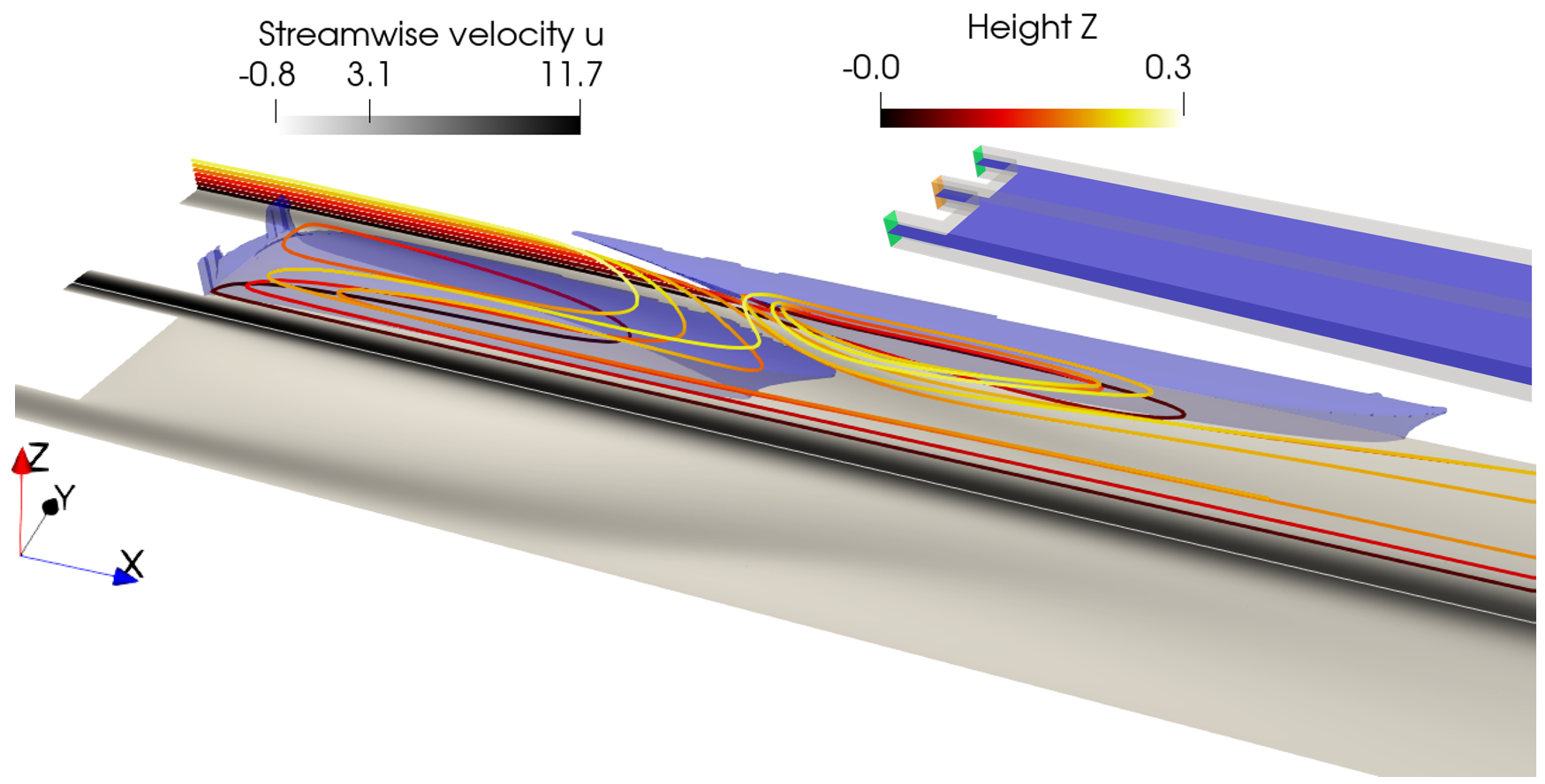}\label{fig:stream 0.7}}\\
    \subfloat[$\Gamma = 1.25, Re = 141$]{\includegraphics[width=.5\linewidth]{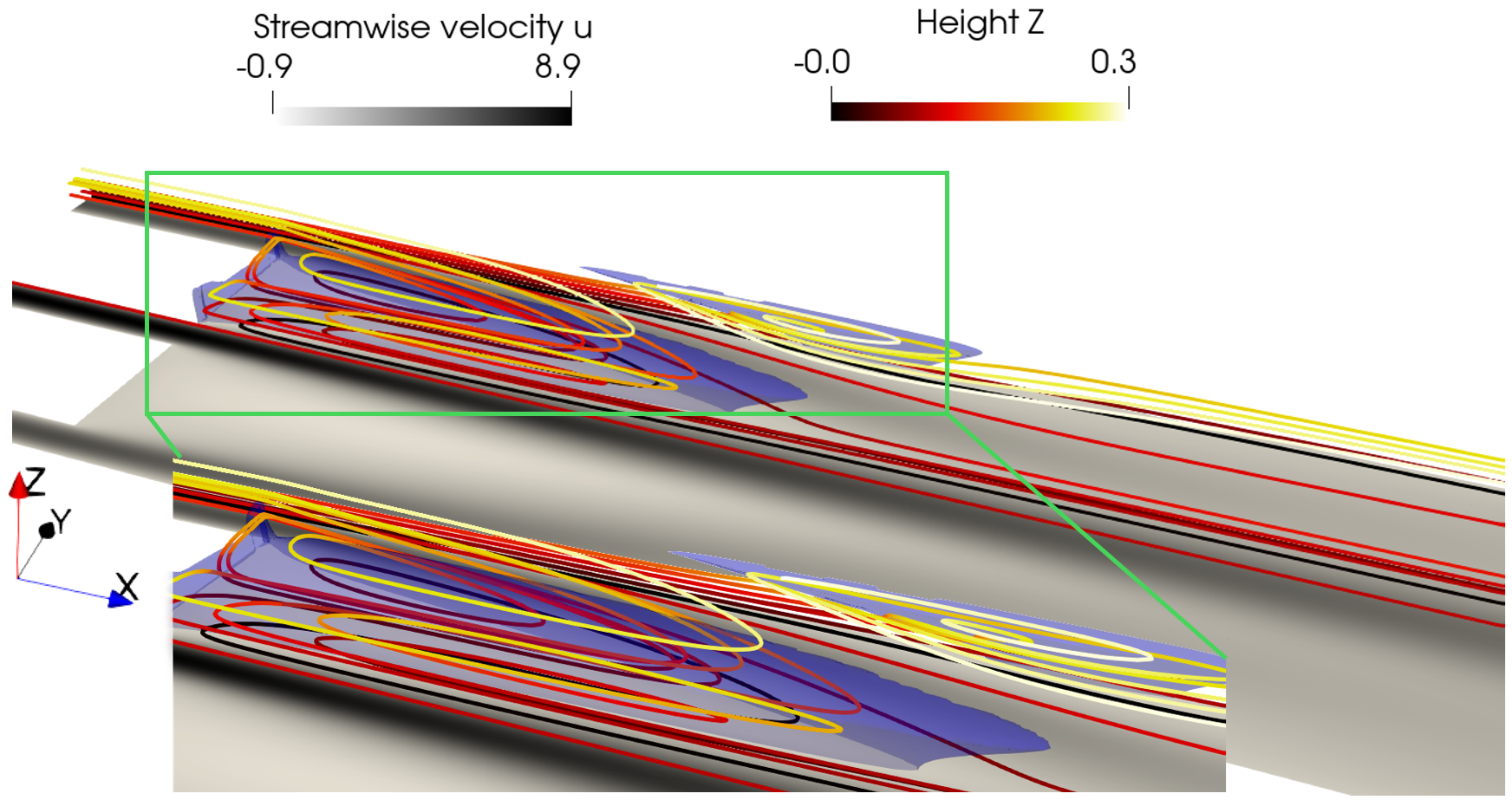}\label{fig:stream 1.25}}
    \subfloat[$\Gamma = 2.5, Re = 86$]{\includegraphics[width=.5\linewidth]{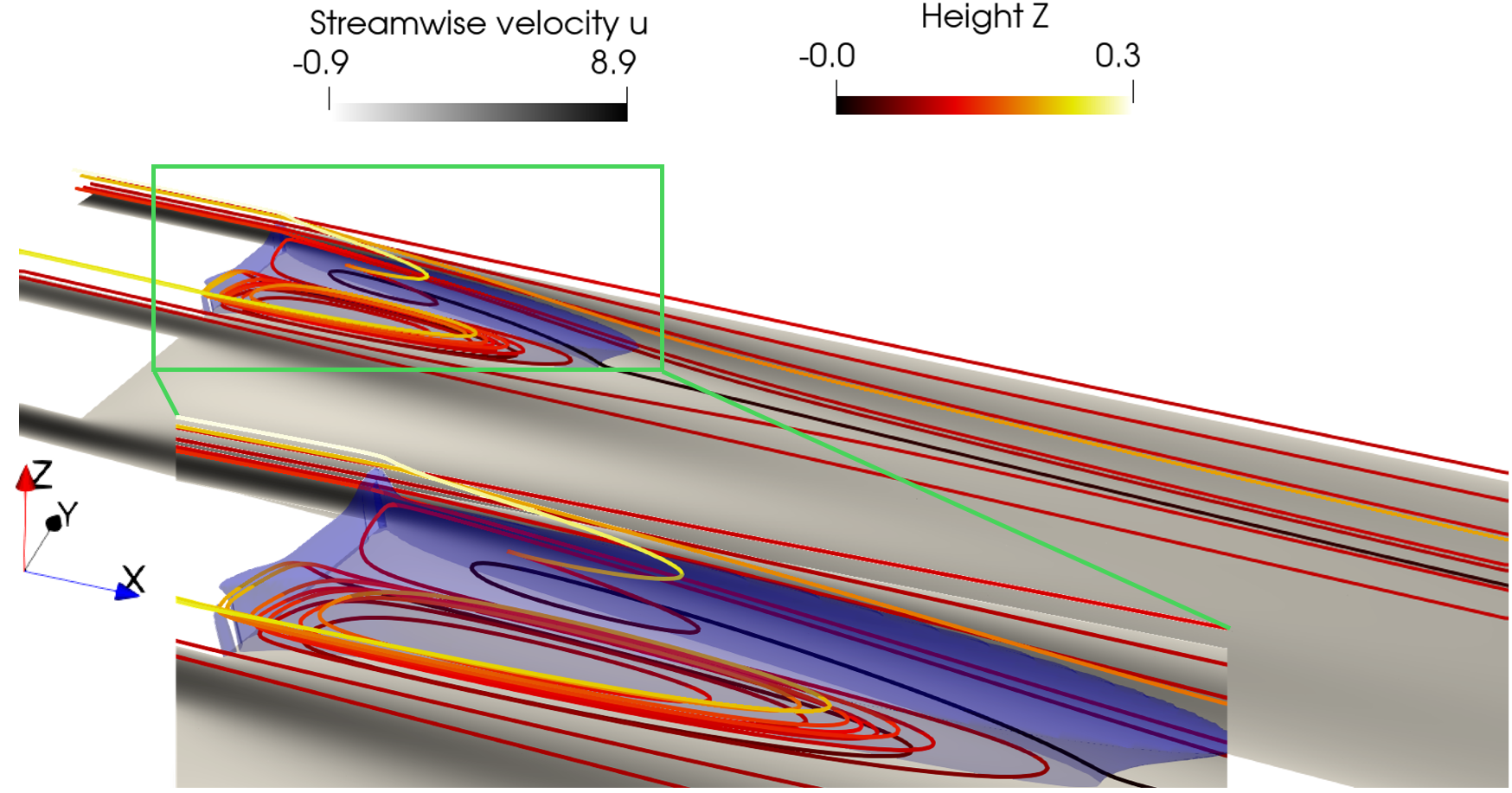}\label{fig:stream 2.5}}
    \caption{Height coloured streamlines, blue translucent iso-contours of null streamwise velocity ($u_x=0$) and mid-plane slice ($z=0$) coloured by the streamwise velocity.}
    \label{fig:Streamlines LSA}
\end{figure}

To understand the 3D nature of the flow at different values of $\Gamma$ a perspective view of streamlines is shown in fig.\ref{fig:Streamlines LSA}, which are coloured by their local height by a heat-map to help visualisation (the mid plane $z=0$ is in black and the walls $z=\tilde{w}/D_h=\frac{3}{2}\tilde{w}\cdot 10^{3}=\pm.3$ are in white). Zones of negative streamwise velocity are made explicit by translucent blue contours of $u_x=0$ on a quarter of the domain ($z>0, y>0$).

In the following description, we use the term "recirculation zone" to refer to a region that appears to contain closed streamlines, with fluid particles seemingly trapped indefinitely. The term "backflow zone" refers to a region where there is backflow in the streamwise direction, but the streamlines do not appear to be closed. In these regions, fluid particles have long residence times but are ultimately advected towards either another backflow zone or the outlet.

As seen in fig.\ref{fig:stream 0.1}, flows of region I are characterised by a single elongated recirculation zone per half domain that spans the whole height of the channel. Three-dimensional effects are negligible due to the scale of the $u_z$ velocity ($10^{-1}$) being smaller than the $u_x$ streamwise velocity ($10$).

When increasing $\Gamma$ to $0.2$ (fig.\ref{fig:stream 0.2}), the recirculation zone elongates and migrates in the downstream direction, while a backflow zone appears right at the inlets. The two are connected by a bridge-like structure that does not span the full height of the channel, near the upper and lower walls at $z=\pm0.3$. The backflow zone near the inlets creates a convergence, bringing the fluid particles towards the mid plane $z=0$, since locally the $u_z$ velocity is of the same scale or larger than the streamwise velocity $u_x$.

At $\Gamma =0.4$ (fig.\ref{fig:stream 0.4}), the recirculation zone at the lateral walls is even more elongated and has split into a recirculation zone downstream and a backflow zone upstream. The recirculation zone downstream exhibits quasi-2D flow, whereas the upstream backflow zone exhibits a strong divergence, transporting fluid particles towards the two opposite top and bottom walls. Once near enough to the wall, they are advected upstream to the backflow zone near the inlets, where they are converged to the mid-plane and finally advected towards the outlet. 

The downstream recirculation zone at the lateral walls, as $\Gamma$ is increased, first detaches from the upstream backflow zone at the side walls, then it is advected towards the outlet and finally disappears, which can be seen at $\Gamma=0.7$ in fig.\ref{fig:stream 0.7}. Apart from the former existence of the recirculation zone at the lateral walls, the flow is similar to what was described for $\Gamma=0.4$ with more elongated backflow zones. 

Eventually, the two backflow zones become disconnected, in the sense that no fluid particle seems to be able to pass through both of them in a single trajectory, as shown at $\Gamma=1.25$ in fig.\ref{fig:stream 1.25}. The upstream backflow zone shrinks as $\Gamma$ increases, disappearing around $\Gamma=2.5$, as shown in fig.\ref{fig:stream 2.5}. 

\subsection{Bifurcation to asymmetric and/or unsteady states}
\subsubsection{Experimental observations}
Past the critical Reynolds number $Re_c$, in zones I and III, the flow bifurcates to a steady asymmetric state which, in the region of parameters explored, is always characterized mainly by a steady deflection of the central jet towards the side walls (we remind the reader that they are located at $\tilde{y}/D_h=\pm1.5$). This deflection should, theoretically, have equiprobability to be oriented to either side; however, due to experimental defects, such as the non-uniformity of the inlets' size (see tab.\ref{tab: dimensions}), we always observed a deflection to the same side. To criticize our results, it should be made explicit that in the case of an already asymmetric channel, the flow has no reason to be symmetric in the first place. However, due to the slight geometric asymmetry of the inlets ($\simeq 1\%$), we expect the system to bifurcate near $Re_c$ through an imperfect pitchfork bifurcation. This means that the asymmetry remains very weak before $Re_c$ and quite suddenly becomes much more important, revealing the topological change associated with the underlying perfect bifurcation as seen in \cref{fig: FRR 01,fig: FRR 125}. 

In zone I, at constant $\Gamma$, the instability materializes by the shrinking of one of the recirculation zones, the expansion of the other one, and a deflection of the jet towards one lateral wall, as seen for $Re\geq 66.3$ in \cref{fig: FRR 01}. Way past $Re_c$ one can see that at $Re =78$ that the jet deflection splits the recirculation region to which it is directed into two. 

In zone III, at constant $\Gamma$, the instability manifests by a shrinking of one of the recirculation zones, the expansion and tilting of the other one, and a deflection of the jet towards the lateral walls, as one can see for $Re\geq 140.3$ in \cref{fig: FRR 125}. 

In zone II, we experimentally observed the presence of "wiggling" pathlines in the sense that on short distances and time scales, the particles start to wiggle around their mean trajectory, resulting in a blurry pathline (see \cref{fig: FRR 07}). At constant $\Gamma$, as the total flow rate $Q$ or the $Re$ increases, the onset of "wiggling" migrates upstream while the wiggling motion seems to increase in amplitude. Despite the presence of these wiggling pathlines, one can distinguish some asymmetry in the length of the backflow zone downstream as well as a deflection of the central jet. 

\subsubsection{Linear Stability analysis (LSA)}

\begin{figure}[H]
    \centering
    \subfloat[]{\includegraphics[width=0.5\linewidth]{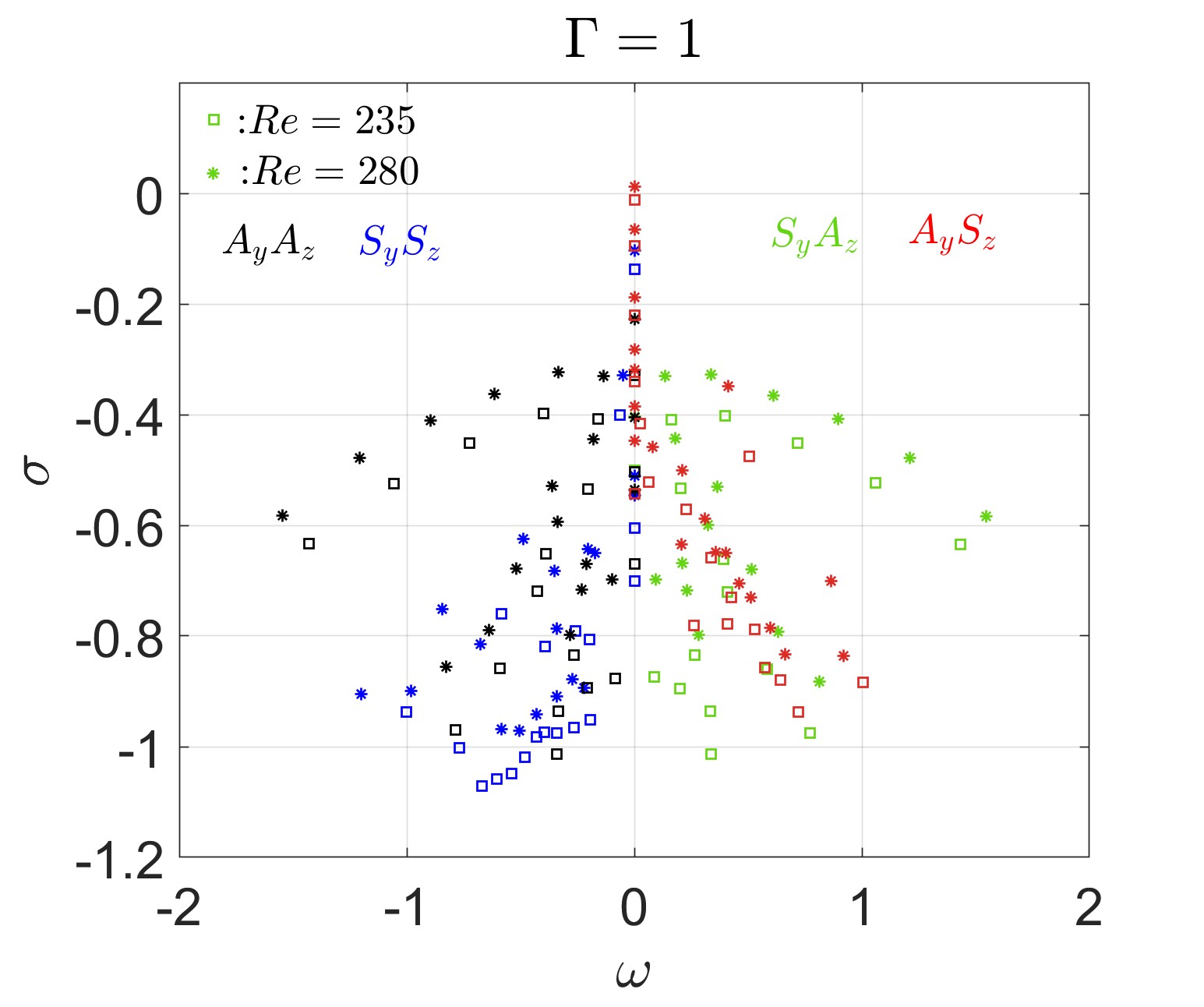}}
    \subfloat[]{\includegraphics[width=0.5\linewidth]{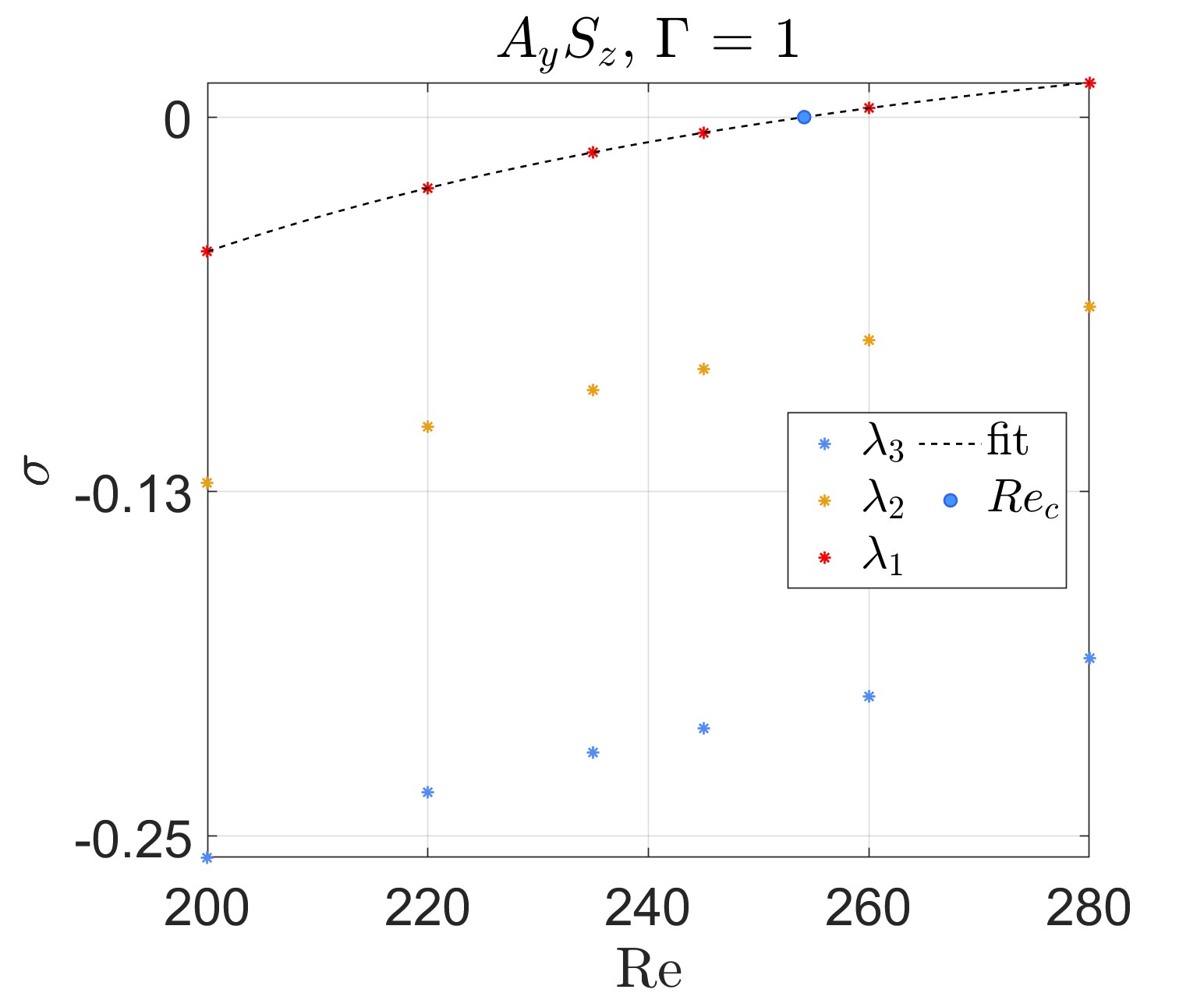}}
    \caption{Example of a linear stability analysis results conducted at $\Gamma=1$. (a) Eigenspectrum of the growth rate $\sigma$ vs the pulsation $\omega$ at $Re=235<Re_c$ (squares) and at $Re=280>Re_c$ (asterisks) for the four families of symmetry breaking, no symmetry breaking ($S_yS_z$) in blue, $y$ and $z$-symmetry breaking ($A_yA_z$) in black, only $z$-symmetry breaking ($S_yA_z$) in green and only $y$-symmetry breaking $A_yS_z$ in red. (b) Growth rate dependence in $Re$ of the three largest growth rates $\sigma_i$. The quadratic fit in $1/Re$ is depicted by a dashed black line, and the LSA predicted value of $Re_c$ by a blue circle.}
    \label{fig:LSA_result_gamma_1}
\end{figure}

The linear stability analysis in zones I and III predicts that the system is unstable with respect to a steady perturbation that breaks only the $y$-symmetry of the base flow. Indeed, we did not find any eigenvalues $\lambda = \sigma + i\omega$ with a positive growth rate $\sigma$ for eigenstates breaking either the $z=0$ plane symmetry or both the $y=0$ and $z=0$ plane symmetries. Figure \ref{fig:LSA_result_gamma_1}a. shows an example of a spectrum plotting the 20 leading eigenvalues for each symmetry breaking at $\Gamma =1$ for two values of $Re$, $Re=235$ below threshold $Re_c$ and $Re=280$ above. As one can see, only the $A_yS_z$ perturbations possess an eigenvalue such that its real part ($\sigma$) becomes positive when the Reynolds number increases beyond threshold. Figure \ref{fig:LSA_result_gamma_1}b. presents the growth rate ($\sigma_i$ depicted with asterisks) of the first three eigenvalues dependence in $Re$ with a quadratic fit in $Re^{-1}$ (depicted by a black dashed line), which gives an estimation of $Re_c$ (depicted by a blue dot). For all values of $\Gamma$ considered for which a threshold is found for $Re<500$, the steady symmetric base state first bifurcates to a steady $y$ asymmetric state, which corresponds to what is found experimentally in regions I and III, at a threshold $Re_c$ close to the experimental one. In region II, the linear stability analysis does not predict an instability threshold for either symmetry-breaking perturbations.

\begin{figure}
    \centering
    \includegraphics[width=\linewidth]{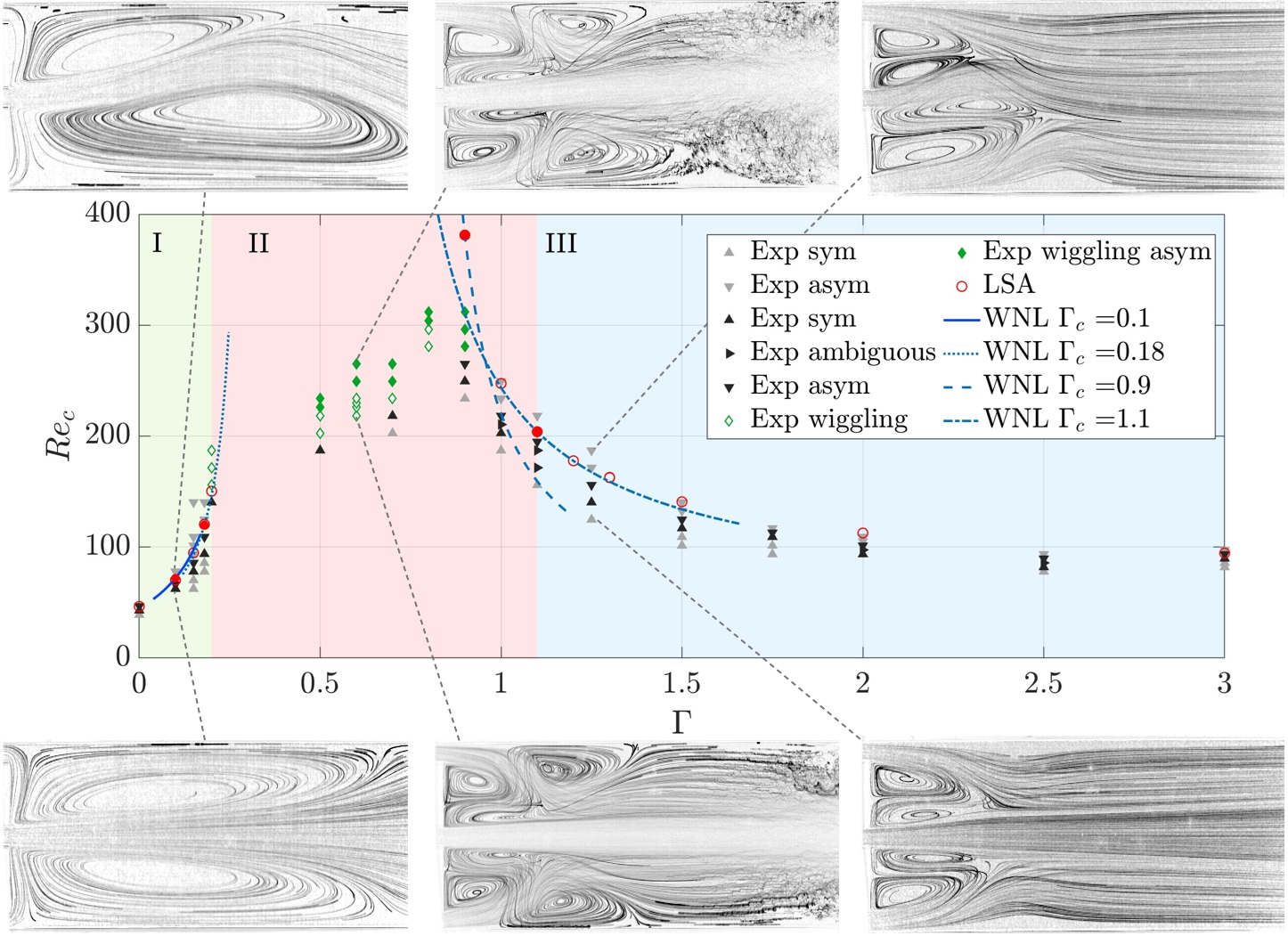}
    \caption{Numerical critical $Re_c$ and experimental data points $Re$ in function of the flow rate ratio $\Gamma$ with weakly nonlinear predicted neutral curves, around points ($\Gamma_c$, $Re_c$) in filled red dots, in blue lines. Experimental measures displaying wiggling pathlines are represented with green diamonds and are filled if they present signs of asymmetry. Measures without wiggling are displayed with grey and black triangles pointing upwards if the flow is symmetric, downwards if the flow is asymmetric, and to the side if the symmetry of the flow is ambiguous. Black triangles are used to show the last(s) measured symmetric flow and the first clear asymmetric one. }
    \label{fig:Rec}
\end{figure}

Figure \ref{fig:Rec} compiles the experimental and numerical results of this paper. First, the three regions that were qualitatively identified by analysing the flow (cf.\ref{section:topology}) are represented by coloured patches (green for region I, red for region II and blue for region III). The experimental observations are depicted by triangles for steady observations and green diamonds for observations with the presence of wiggling pathlines. Black triangles are used to depict at each $\Gamma$ the highest $Re$ where symmetric flow is observed and the lowest $Re$ at which asymmetry is observed, while grey triangles depict additional measurements (above and below the experimental $Re_c$). Upward-facing triangles stand for steady symmetric observations (see left/right pictures below), downward-facing ones for steady asymmetric (see left/right pictures above) and side-facing triangles for ambiguous symmetry observations. Finally, empty green diamonds denote experimental points where wiggling pathlines are observed under a mean symmetric flow (see middle picture below), whereas filled green diamonds depict an asymmetric mean flow (see middle picture above). One can observe a gap in the experimental results in the region $\Gamma\in[0.2,0.4]$ because the recirculation region is longer than the field of view, preventing assessment of its symmetry. The predicted thresholds $Re_c$ by the linear stability analysis for the first bifurcation, which is a steady $y-$symmetry breaking, are denoted with empty red circles. Finally, weakly nonlinear approximate neutral curves, shown in blue dashed/dotted/dashed-dotted/solid, are computed around 4 different $(Re_c,\Gamma_c)$ pairs, shown with filled red circles  (only the values of $\Gamma_c$ are given in the legend, but the associated $Re_c$ can be read from the figure).

In regions I and III, experimental and linear stability analysis $Re_c$ show a good agreement, with an increasing discrepancy in the values of $Re_c$ when approaching region II. Moreover, as $\Gamma$ is near either side of region II, the experimental and numerical thresholds for the appearance of an asymmetric flow rapidly increase. However, in region II, while experiments show the appearance of wiggling pathlines on a mean symmetric state and at a higher $Re$, the existence of wiggling pathlines on a visually mean asymmetric state, the linear stability analysis does not provide a critical value of $Re$, which can be seen from the absence of red circles.

The weakly nonlinear analysis (WNL) developed in \cref{sub:WNL} not only characterizes the nature of the bifurcation in terms of $Re$ (supercritical or subcritical), but also yields an approximation of the growth-rate dependence in $Re$ and $\Gamma$, together with a prediction of the imperfect bifurcation when an antisymmetry of the flow rate $\varepsilon^3\phi$ is considered.

The approximate neutral curves, in the neighbourhood of their construction, follow the predicted $Re_c$ and provide a satisfactory prediction of $\sigma(Re,\Gamma)$ (\cref{fig:pred WNL 1.1} in the appendix). On both sides of region II, the predicted $Re_c$ increases sharply, confirming the strong re-stabilisation of the steady $y$-symmetry breaking bifurcation in this region, in accordance with the quenching of this instability for $\Gamma\in]0.2, 0.9[$ (region II). 

\begin{figure}[H]
    \centering
    \includegraphics[width=1\linewidth]{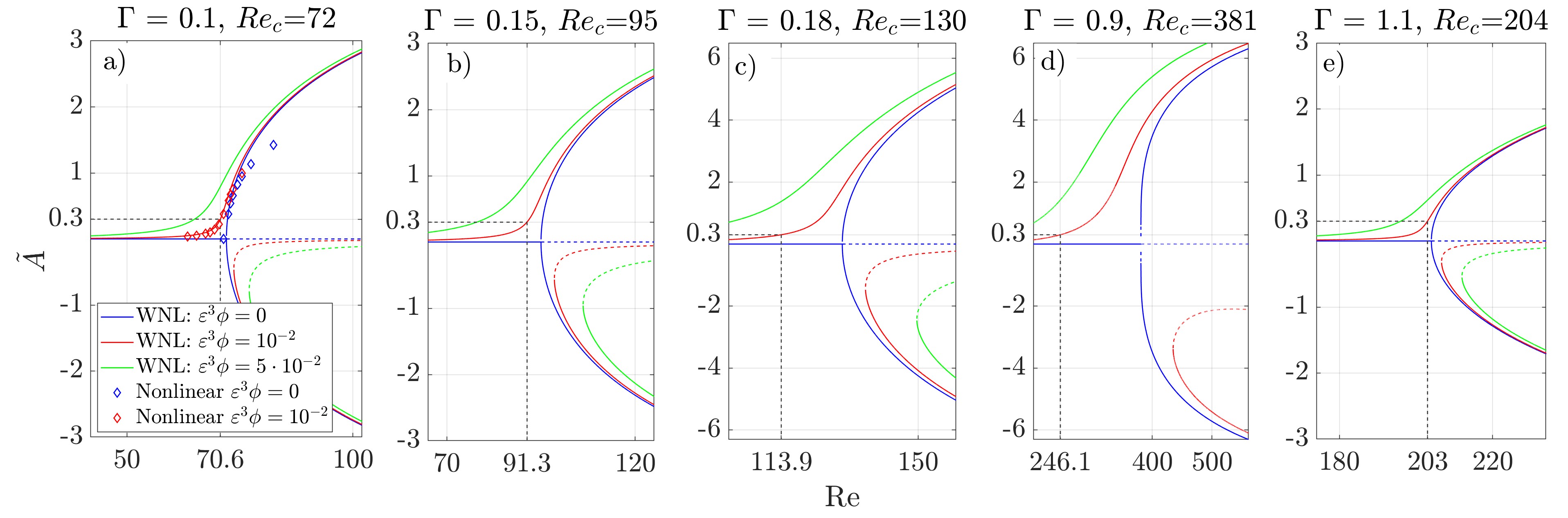}
    \caption{Comparison between the WNL-predicted perfect bifurcation scenario in blue and imperfect bifurcation scenarios with a flow rate antisymmetry $\varepsilon^3\phi=10^{-2}$ in red and $\varepsilon^3\phi =5\cdot10^{-2}$ in light green. The amplitude $a$ has been rescaled by a factor $||\vec u_1||/\langle \vec u_1,\udvec \rangle$ to show the saturation amplitude $\tilde A$ of the eigenmode $\vec u_1$ with unitary norm. In the first panel, the fully nonlinear steady numerical solution with the full domain simulated is displayed with blue/orange diamonds. Note that panels a)-c) and e) are plotted using \cref{eq: amp_eq_a} while panel d) is plotted using the $5^{th}$ order amplitude equation \cref{eq:5thorder} in appendix C.} 
    \label{fig:asym}
\end{figure}

Using the derived WNL coefficients, one can infer the nature of the bifurcation by examining the sign of the ratio $\lambda/\mu$ (note that both are real since the bifurcation is steady). This ratio is found to be negative for all the $(Re_c,\Gamma_c)$ computed except at $\Gamma_c =0.9$, $Re_c = 381$ as shown in \cref{tab: amp_eq}. This means that the symmetry-breaking bifurcation is a supercritical pitchfork bifurcation for all values of $\Gamma$ but $\Gamma=0.9$, where it is subcritical. However, the subcritical behaviour at $\Gamma =0.9$ is not very pronounced as the hysteretic region is very narrow $Re_c^* - Re_c=0.025$ (see \cref{fig:asym}.d).

To take into account the experimental reality and the asymmetry of the channel reported in \cref{tab: dimensions}, we model this by considering a flow rate antisymmetry of $\varepsilon^3\phi=10^{-2}$. Keeping the lowest order terms such that the amplitude equation has a physical meaning, we obtain the perfect (blue lines) and imperfect (red lines) bifurcations presented in \cref{fig:asym}, where dashed lines represent unstable solutions and solid ones stable solutions. Note that the amplitude ranges across the different panels are not the same. To validate our model, we performed several fully steady nonlinear simulations at $\Gamma =0.1$, lifting the symmetry conditions by considering the full domain and the asymmetric inlet flow condition $u|_{\Omega_{l_\pm}} = (1\pm \varepsilon^3\phi)C^{-1}_{l_\pm}u_{in}(y,z;y_{l},w,H)$. Taking the scalar product between the velocity field $\vec u$, obtained with this asymmetric inlet fluxes but a symmetric geometry, and the adjoint mode $\udvec$ (calculated in the absence of imperfection in the WNL analysis), we obtain the saturated amplitude $a=\langle\udvec, \uv{}{}{}\rangle/\langle\udvec,  \uv{1}{}{}\rangle$ of the normalized perturbation. Comparing the effect of a given asymmetry strength $\phi$ at different $\Gamma$, one can see that near the local stabilisation of this perturbation, in zone II, the sensitivity to inlet flow rate asymmetry increases. Due to the experimental method for determining the critical Reynolds one could imagine that the naked eye could only perceive an asymmetry when the perturbation norm is larger than a certain threshold, which we arbitrarily set for instance to $0.3$, which would result in an experimental critical Reynolds lower than the numerical one, as seen in \cref{tab:rescaled Rec,fig:asym}, which could explain the
increase in discrepancy between the experimentally measured $Re_{c,exp}$ and the one predicted from LSA $Re_c$, when approaching both boundaries of region II from regions I and III. 

\begin{table}[H]
    \centering
    \begin{tabular}{|c|c|c|c|}\hline
         $\Gamma$ & $Re_c$ & $Re^{0.3}_c$& $Re^{exp}_c$ \\\hline
          0.1 & 72 & 70.6 & $64\pm2$ \\
          0.15 & 95 & 91.3& $81.5\pm 3.5$ \\
          0.18 & 130 & 113.9  & $101.5\pm7.5$ \\
          0.9 & 381  & 246.1 & $257.5\pm 7.5$ \\
          1.1 & 204 & 203 &  $182\pm11$\\\hline
    \end{tabular}
    \caption{Numerical critical Reynolds $Re_c$, WNL critical Reynolds $Re_c^{0.3}$ using the criterium $\tilde A=0.3$ and experimental Reynolds $Re_c^{exp}$.}
    \label{tab:rescaled Rec}
\end{table}

\subsubsection{ Temporal direct numerical simulations (T-DNS)}
To investigate the physical mechanism active in region II, where LSA fails to predict a threshold for instability, while experimentally one can see blurry pathlines that could either be an indicator of an oscillating growing perturbation or a response to external/experimental noise, we performed two T-DNS with the real (asymmetric) geometry of the channel using the spectral code Nek5000 as reported in Section~\ref{sub:base_state}. A T-DNS was carried out at $Re=90$, approximately $20\%$ above the numerical critical Reynolds number $Re_c=72$ for $\Gamma =0.1$ to verify if spontaneous symmetry breaking was observed without artificial noise forcing. A second DNS was performed in region II, above the experimental Reynolds number at which blurry pathlines are first observed, at $\Gamma = 0.5$ and $ Re = 300$ to elucidate the origin of the experimentally observed fluctuations.

\begin{figure}[H]
    \centering
    \includegraphics[width=1\linewidth]{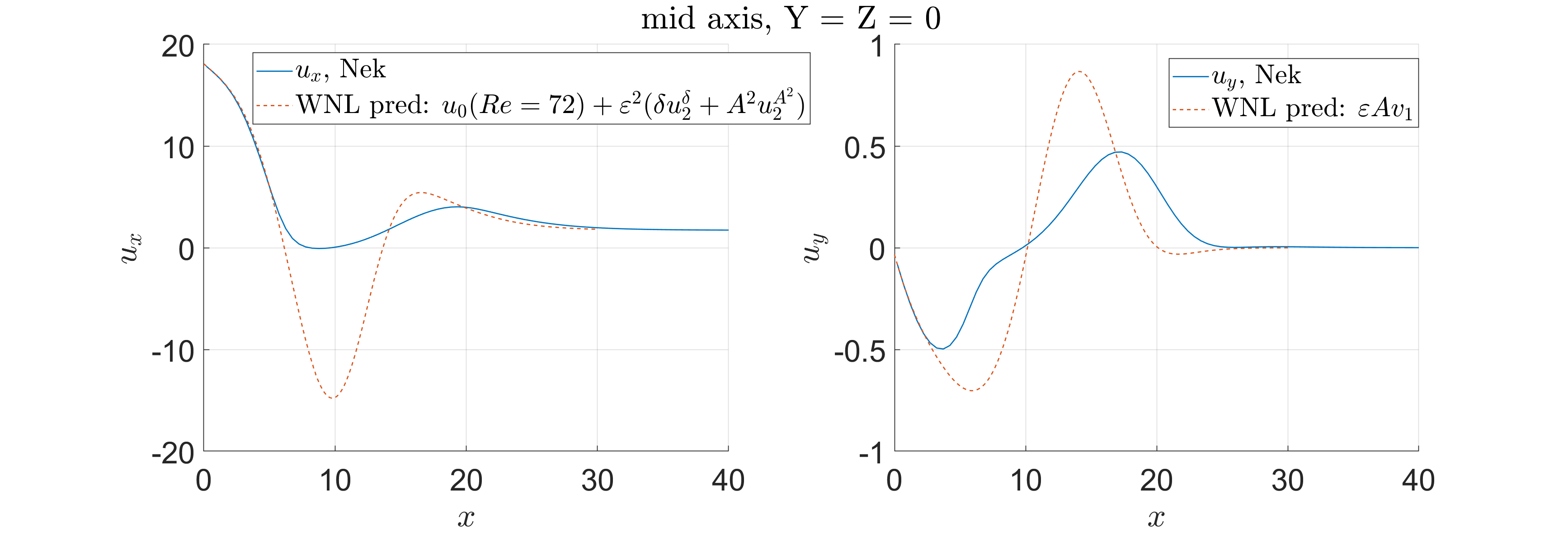}
    \caption{Comparison of the total velocity $\vec u$ computed in Nek500, in blue solid line, and the WNL predicted one, in orange dashed line, at the mid axis $Y=Z=0$. }
    \label{fig:comp_Nek_WNL}
\end{figure}

At $\Gamma =0.1$, the system converges, with an $L^2$ norm of the absolute residual less than $10^{-7}$, to an asymmetric steady solution governed by the same bifurcation scenario that was observed experimentally and numerically (LSA and stationary DNS), where the central jet breaks only the $y$-symmetry plane, being deflected to one of the side walls with an elongation of one of the recirculation zones and a shrinking of the other one. Given the substantial difference between the critical Reynolds predicted by LSA ($Re_c(\Gamma =0.1)=72$) and the Reynolds number simulated ($Re=90$), the non-linearities and the amplitude of the eigenmode cannot be compared with the WNL predictions (see \cref{fig:comp_Nek_WNL}), with almost quantitative agreement. At $\Gamma = 0.5$, the system also converges, with an $L^2$ norm of the residuals inferior to $10^{-6}$, to a steady solution. This solution possesses an asymmetry in the y-plane $\max\limits_x(u_y|_{y=z=0})=4\cdot10^{-3}$ that is of comparable order to the asymmetry of the channel, discarding the hypothesis of a missed stable solution by the LSA conducted. This suggests that the observed fluctuations, materialised by the wiggling trajectories of tracers, likely result from noise amplification mechanism, where the noise is most likely generated by the syringe pumps or pressure controller (see appendix \ref{appendix: prelim}), the camera ventilation or other sources of vibrations.

\section{Conclusion}

In the present paper, we investigated the flow in a three-dimensional confined sudden expansion with lateral inflow with two planes of symmetry at low Reynolds number, $Re<300$, and lateral to central inlet flow rate ratio, $\Gamma\in[0, 3]$. First, we sought to characterise the topology change the flow undergoes as the flow rate ratio increases by qualitatively identifying recirculation regions and backflow zones, thereby showing the existence of 3 distinct flow regimes. We demonstrated experimentally that at low $\Gamma \in [0, 0.15]$ and high $\Gamma\in[0.9, 3]$, there exists a critical value of the Reynolds number, $Re_c$, below which the flow is symmetric, while above, the flow displays a steady asymmetry characterised by a steady deflection of the central jet to either of the side walls. In the intermediate region $\Gamma\in[0.4,0.9]$, the experiments show the emergence of fluctuations that remain poorly understood and would require more detailed experimental and theoretical investigations. 

Performing a linear stability analysis with a finite element method, we showed that the experimentally observed steady asymmetry arises from a ($y$-)symmetry-breaking steady instability, while no unstable modes breaking other spatial symmetries and time-invariance could be identified. Using a weakly nonlinear approach, we derived an amplitude equation to predict the dependence of growth rates in $Re$ and $\Gamma$, which shows that the intermediate region II $\Gamma\in[0.2, 0.9]$ is a region of stabilisation where the steady ($y$-) symmetry-breaking instability does not have a critical Reynolds. Additionally, we showed that this bifurcation is a supercritical pitchfork bifurcation for almost all values of $\Gamma$ considered, where the instability is found to exist ($\Gamma \notin [0.2,0.9]$). Temporal direct numerical simulations with a spectral element code (Nek5000), which took into account the geometrical asymmetry, did not show any evidence of an instability that was missed by the linear stability of the symmetric channel in region II. The T-DNS indeed showed an asymmetry in the longitudinal component of the velocity of the order of the flow-rate asymmetry imposed, failing to present a significant amplification.

Further work should be done to investigate the physical mechanisms at play in this configuration. In the region of low $\Gamma<0.15$, we expect a physical mechanism akin to the one at play in planar sudden expansion, which is stabilised by the confinement transverse to the flow and the presence of the two lateral inlets, which at low $\Gamma$, only changes the flow topology slightly. In the region of intermediate $0.15<\Gamma<0.9$, we believe that more quantitative experiments should be led to verify and understand the flow unsteadiness and behaviour using $\mu$-PIV or PTV, which might prove to be difficult due to the three-dimensionality of the flow. Numerical investigation of the system response to noise or forcing using either a resolvent approach or instationary direct numerical simulations might be of interest to understand the discrepancy between our experiments and linear stability analysis. The dominance of a steady bifurcation in region III, where the inner jet is weak and a confined wake flow forms, should also be understood, given the existing predictions in the literature. In purely 2D flows, an intermediate level of confinement is known to promote the absolute instability of wakes \cite{Juniper2011,Biancofiore2011}, resulting in an oscillatory global instability forming an alternated vortex street that should possibly compete with the steady bifurcation identified in this work.  

\medskip

\bibliography{sample}

\newpage
\appendix
\section{Preliminary study} \label{appendix: prelim}

\begin{figure}[H]
    \centering
    \subfloat[Schematic representation of the preliminary geometry used.]{\includegraphics[width=0.2\linewidth]{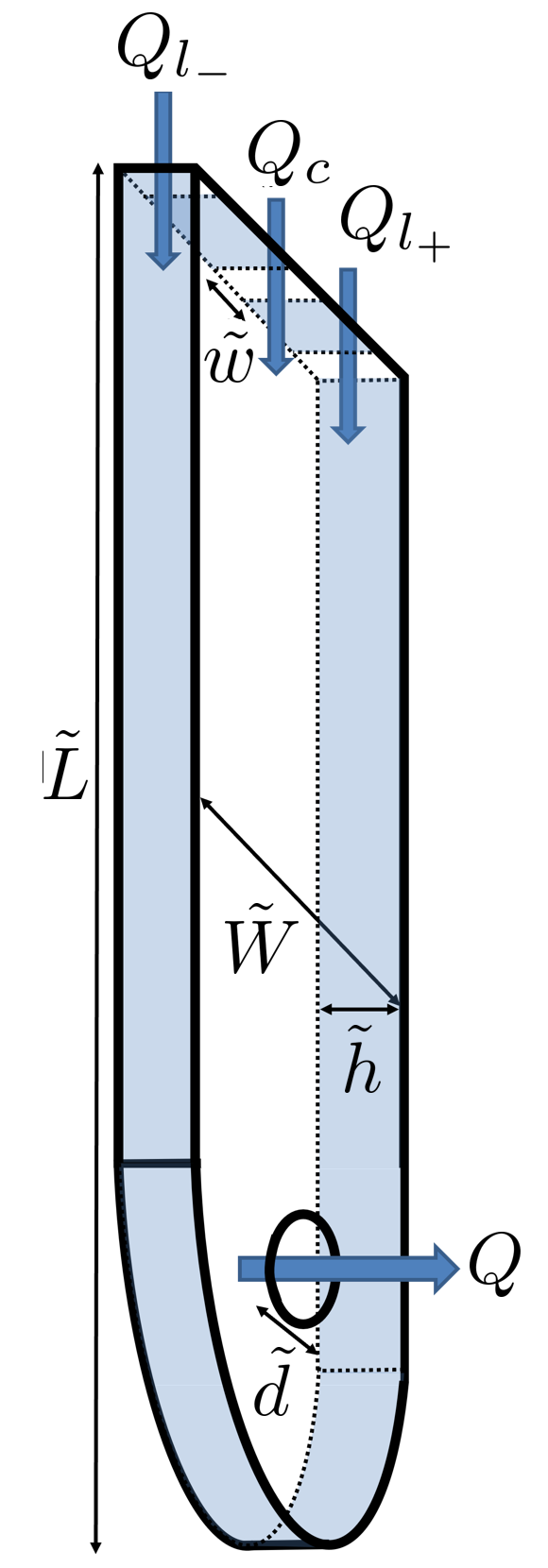}\label{fig:prev_geom}}\hspace{1cm}
    \subfloat[Comparison of the preliminary results obtained using geometry of \cref{fig:prev_geom} and the syringe pump (blue symbols) or the pressure controller (orange symbols) and the present experimental results (in green diamonds and black circles). Observation of wiggling is depicted by diamonds while observation of a steady asymmetric bifurcation is denoted by circles.]{\includegraphics[width=0.6\linewidth]{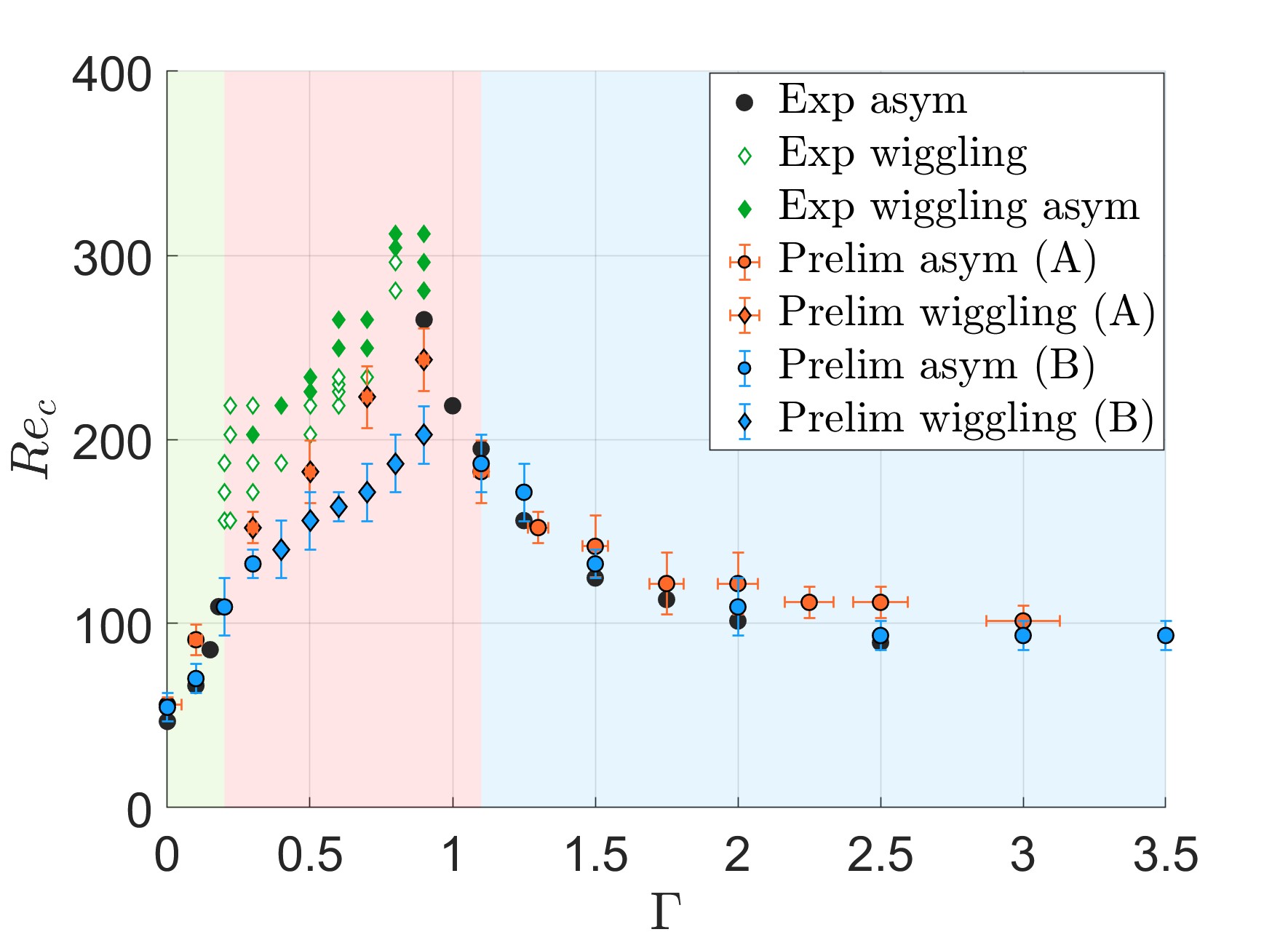}\label{fig:comp_prelim_rec}}
    \caption{Preliminary geometry (a) and associated bifurcation results compared to results obtained with the channel studied in the main text (b).}
\end{figure}

In a preliminary study to investigate this triple inlet channel configuration experimentally, we used a different geometry, which is depicted in \cref{fig:prev_geom}. The dimensions of the channel are the same as those considered in this paper, with the exception of the length of the channel $\tilde L=25$ mm and the out-of-plane outlet of diameter $\tilde d=0.4$ mm, which is linked to a reservoir through micro-tubing (instead of an open outflow). To obtain the same operating conditions in both the lateral inlet, a 90-degree split was printed inside the chip instead of the smooth one designed and used throughout this paper. Apart from the geometry, the experimental visualisation setup and apparatus were the same. We used to different strategies for fluid injection inside of the channel: (i) the syringe pump setup previously discussed (denoted (A) in orange symbols in \cref{fig:comp_prelim_rec}) and (ii) a Fluigent high pressure source (denoted (B) in blue symbols in \cref{fig:comp_prelim_rec}) with a monitoring of the central flow rate $Q_c$ using a Bronkhorst mini-Coriflow flow-meter and of the total lateral flow rates $Q_{l}=Q_{l_+}+Q_{l_-}$ inferred from the measurement of the outlet flow rate $Q$ using a Sartorius CP12001S Weighing system. 

The experimental thresholds in regions I and III are quite comparable to those obtained with the carefully designed chip described in the main text, while in region II, where numerics/theory and our present experiments are not in agreement is less robust. While we believe that the changes in geometry brought to this preliminary channel design were important to ensure proper comparison between experiments and numerics, they did not fully resolve the pre-existing discrepancy in the transition toward wiggling and symmetry breaking. However, the existence of such a transition for both geometries and fluid injections (pressure controller/syringe pump) supports the hypothesis of a transition of the flow as a consequence of the amplification of external disturbances. 

\section{Amplitude equations}\label{annexe:amp_eq}
At $1^{st}$ order, we found that the first bifurcation to an asymmetric flow is due to a steady mode $\qv{1}{}{}=\qv{1}{1}{}$ at $Re=Re_c(\Gamma)$. In other words: 
\begin{align*}
    &\hat L\, \qv{1}{j}{} = (\underbrace{\sigma_j}_{<0,\,\, \forall j\neq1}+i\omega_j) \hat B \qv{1}{j}{},\\
    &\hat L \qv{1}{1}{}=(\underbrace{\sigma}_{\overset{.}{=}\,0\text{ at } Re_c} + i\underbrace{\omega}_{\text{steady mode}}) \hat B\qv{1}{1}{} = 0.
\end{align*}

We also constructed an adjoint operator $\hat L\dd$ with respect to the Hermitian scalar product ($\langle \vec a, \vec b \rangle = \sum\limits_{i=1}^{4}\int_\Omega \vec{a}^{*,T}\vec b \, dx$) and found an adjoint mode $\qdvec$ such that: 

\begin{equation*}
    \hat L\dd \qdvec= 0, \qquad \langle\qdvec, \qv{1}{j}{}\rangle = C_j \delta_{i,j} .
\end{equation*}

Therefore, at $2^{nd}$ order, using the linearity of the operator and hence the solution, we obtain the following equations:

\begin{align}
    &\qquad \qquad\qv{2}{}{} = A^2\qv{2}{A^2}{}+ \delta\qv{2}{\delta}{}+ f\qv{2}{f}{},\quad \qv{2}{i}{}\in{S_yS_z}\\
    &\hat{L} \qv{2}{A^2}{} = -C(\uv{1}{}{},\uv{1}{}{}), \qquad \text{with homogeneous BC on }\partial\Omega,\\
    &\hat{L} \qv{2}{\delta}{} = -\Delta\uv{0}{}{}, \qquad\qquad\,\,\,\,\,\text{with homogeneous BC on }\partial\Omega,\\
    &\hat{L} \qv{2}{f}{} = 0, \qquad\uv{2}{f}{}=\begin{cases} -hWp(y,z)\quad \text{   on }\partial\Omega_{c},\\ \frac{hW}{2}p(y,z) \quad\text{   on }\partial\Omega_{l}, \end{cases}\\
    &\qquad\qquad \text{and homogeneous BC on }\partial\Omega\backslash \partial\Omega_i.
\end{align}

Where we recall that $\varepsilon^2\delta=Re_c^{-1} - Re^{-1}$ is a measure of distance to threshold in the Reynolds number and $\varepsilon^2 f = (1+\Gamma_c)^{-1}-(1+\Gamma)^{-1}$ is a measure of distance to threshold in the flow rate ratio $\Gamma$.

All of these equations are easily invertible as $\qv{2}{i}{}$ all belong to $S_yS_z$, which is not in the kernel of the operator $\hat{L}$. To give a bit of intuition, $\qv{2}{A^2}{}$ is a mean-flow correction due to the growing pertubation $\varepsilon A(T)\qv{1}{}{}$, $\qv{2}{\delta}{}$ is a mean flow correction to take into account the evolution of the flow with $Re$ around $Re_c$ and $\qv{2}{f}{}$ to take into account evolution of the flow with changes in $\Gamma$.

Once the second-order fields are computed, we find the following bulk equations and boundary conditions at $3^{rd}$ order:

\begin{align}
    &\hat{L} \qv{3}{}{} = -\partial_TA\uv{1}{}{}-A^3C(\uv{2}{A^2}{},\uv{1}{}{})-A\left[\delta \left( C(\uv{2}{\delta}{},\uv{1}{}{}) +\Delta\uv{1}{}{}\right) + f C(\uv{1}{}{},\uv{2}{f}{}) \right], \label{annexe: eq uv3}\\
    & \qquad \qquad \uv{3}{}{}=\begin{cases} 0\quad \text{   on }\partial\Omega_{c},\\
    \phi\,\,\text{sign}(y) \dfrac{hW\Gamma }{2(1+\Gamma)}u_{in}(y,z) \quad\text{   on }\partial\Omega_{l}, \end{cases}\\
    & \qquad \qquad \text{ and homogeneous BC on }\partial\Omega\backslash \partial\Omega_i, \qv{3}{}{}\in A_yS_z.
\end{align}

At this point, equations (25-27) is potentially non-physical. Indeed, the forcing term, or the right-hand side of the equation, has the same symmetries as $\uv{1}{}{}$ (and pulsation $\omega=0$), therefore forcing the system at resonance. However, using the adjoint mode $\qdvec$ of $\qv{1}{}{}$, one can prevent this resonance by projecting the forcing on the orthogonal plane to $\qv{1}{}{}$. This method is often known as the Fredholm alternative/solvability condition/orthogonality condition:

\begin{align}
    &\Longleftrightarrow  0\overset{.}{=}\langle \hat{L}\dd \qdvec, \uv{3}{}{} \rangle = \langle \qdvec, \hat{L}\uv{3}{}{}\rangle + \langle T\dd\cdot \vec n, \uv{3}{}{}\rangle_{\partial \Omega_{l}} \\
    &\qquad \quad = \langle\qdvec, \vec F_3\rangle +\phi\,\, \underbrace{\text{sign}(y) \dfrac{hW\Gamma }{2(1+\Gamma)} \langle T\dd\cdot \vec n, u_{in}(y,z) \vec e_x\rangle_{\partial \Omega_{l}}}_{\theta\langle\qdvec, \qv{1}{}{}\rangle}\\
    &\Longrightarrow \partial_T A \langle\qdvec, \qv{1}{}{}\rangle = - A^3\underbrace{\langle\qdvec, C(\uv{2}{A^2}{},\uv{1}{}{})\rangle}_{\mu \langle\qdvec, \qv{1}{}{}\rangle} - A\left[\delta\underbrace{\langle\qdvec, C(\uv{2}{\delta}{},\uv{1}{}{}) + \Delta\uv{1}{}{}\rangle}_{\lambda \langle\qdvec, \qv{1}{}{}\rangle} + f \underbrace{\langle\qdvec, C(\uv{2}{f}{},\uv{1}{}{})\rangle}_{\alpha \langle\qdvec, \qv{1}{}{}\rangle}\right] + \phi \theta \label{annexe: amp_eq}
\end{align}

Once the coefficients $\lambda$, $\mu$, $\alpha$, $\theta$, found one can replace $\partial_T A$ in eq. \ref{annexe: eq uv3} by its expression in eq. \ref{annexe: amp_eq} and solve for $\qv{3}{}{} = A^3\qv{3}{A^3}{}+A\delta\qv{3}{A}{\delta}+Af\qv{3}{A}{f}+\phi\qv{3}{\phi}{}$ as the forcing is now orthogonal to $\uv{1}{}{}$. 
One can then go at $4^{th}$ order:

\begin{align}
    \hat{L} \qv{4}{}{} =& -\partial_T(A^2)\uv{2}{A^2}{}-A^4\left[C(\uv{2}{A^2}{},\uv{2}{A^2}{})/2 + C(\uv{3}{A^3}{},\uv{1}{}{})\right]\nonumber\\
    &-A^2\left[\delta \left( C(\uv{2}{\delta}{},\uv{2}{A^2}{})+ C(\uv{3}{\delta}{A},\uv{1}{}{}) +\Delta\uv{2}{A^2}{} \right) + f\left(C(\uv{2}{f}{},\uv{2}{A^2}{})+ C(\uv{3}{f}{A},\uv{1}{}{})\right) \right]\nonumber\\
    & - A\phi C(\uv{3}{\phi}{},\uv{1}{}{}) -\delta^2 \left(C(\uv{2}{\delta}{},\uv{2}{\delta}{})/2 +\Delta\uv{2}{\delta}{}\right) -f^2 C(\uv{2}{f}{},\uv{2}{f}{})/2-f\delta C(\uv{2}{f}{},\uv{2}{\delta}{})\\ & \qquad \text{ and homogeneous BC on } \partial \Omega, \qquad \qv{4}{}{}\in S_yS_z\nonumber \\
    & \hat{L} \qv{4}{A^4}{} = -\left[C(\uv{2}{A^2}{},\uv{2}{A^2}{})/2 - C(\uv{3}{A^3}{},\uv{1}{}{})\right] -2\mu \uv{2}{A^2}{}\\
    & \hat{L} \qv{4}{A^2\delta}{} = -C(\uv{2}{\delta}{},\uv{2}{A^2}{})- C(\uv{3}{\delta}{A},\uv{1}{}{}) -\Delta\uv{2}{A^2}{} -2\lambda \uv{2}{A^2}{}-2\mu\uv{2}{\delta}{}\\
    & \hat{L} \qv{4}{A^2f}{} = -C(\uv{2}{f}{},\uv{2}{A^2}{})- C(\uv{3}{f}{A},\uv{1}{}{}) -2\alpha \uv{2}{A^2}{}-2\mu\uv{2}{f}{}\\
    & \hat{L} \qv{4}{A \phi}{} = -C(\uv{3}{\phi}{},\uv{1}{}{}) -2\theta \uv{2}{A^2}{}\\
    & \hat{L} \qv{4}{\delta^2}{} = -C(\uv{2}{\delta}{},\uv{2}{\delta}{})/2 -\Delta\uv{2}{\delta}{}-2\lambda\uv{2}{\delta}{}\\
    & \hat{L} \qv{4}{f^2}{} = -C(\uv{2}{f}{},\uv{2}{f}{})/2-2\alpha \uv{2}{f}{}\quad\\
    & \hat{L} \qv{4}{f\delta}{} = -C(\uv{2}{f}{},\uv{2}{\delta}{})- 2\alpha\uv{2}{\delta}{}-2\lambda\uv{2}{f}{}
\end{align}

And finally at $5^{th}$ order:

\begin{align}
    \hat{L} \qv{5}{}{} =& -\partial_T(A^3)\uv{3}{A^3}{}-\partial_TA\left[\delta\uv{3}{A\delta}{} +f\uv{3}{f}{}\right]-\partial_{T_2}A \uv{1}{}{}\nonumber\\
    &-A^5\underbrace{\left[C(\uv{2}{A^2}{},\uv{3}{A^3}{}) + C(\uv{4}{A^4}{},\uv{1}{}{})\right]}_{\gamma\langle \qdvec, \qv{1}{}{}\rangle}\nonumber\\
    &-A^3\Biggl[\delta \underbrace{\left( C(\uv{2}{\delta}{},\uv{3}{A^3}{})+ C(\uv{3}{\delta}{A},\uv{2}{A^2}{})+ C(\uv{4}{\delta}{A^2},\uv{1}{}{}) +\Delta\uv{3}{A^3}{} \right)}_{\mu_2^\delta\langle \qdvec, \qv{1}{}{}\rangle} \nonumber\\
    & \qquad+ f\underbrace{\left(C(\uv{2}{f}{},\uv{3}{A^3}{})+ C(\uv{3}{f}{A},\uv{2}{A^2}{})+ C(\uv{4}{fA^2}{},\uv{1}{}{})\right)}_{\mu_2^f\langle \qdvec, \qv{1}{}{}\rangle} \Biggr]\nonumber\\
\end{align}
\begin{align}
    & - A^2\phi \underbrace{\left[C(\uv{3}{\phi}{},\uv{2}{A^2}{}) + C(\uv{4}{A\phi}{},\uv{1}{}{}) \right]}_{\chi\langle \qdvec, \qv{1}{}{}\rangle}\nonumber\\
    &-A\delta^2 \underbrace{\left[C(\uv{2}{\delta}{},\uv{3}{A\delta}{})+C(\uv{4}{\delta^2}{},\uv{1}{}{}) +\Delta\uv{3}{A\delta}{}\right]}_{\lambda^\delta_2\langle \qdvec, \qv{1}{}{}\rangle}\nonumber\\
    &-Af^2 \underbrace{\left[C(\uv{2}{f}{},\uv{3}{Af}{})+C(\uv{4}{f^2}{},\uv{1}{}{})\right]}_{\alpha^f_2\langle \qdvec, \qv{1}{}{}\rangle}\nonumber\\
    &-A\delta f \underbrace{\left[C(\uv{4}{f\delta}{},\uv{1}{}{})+C(\uv{3}{A\delta}{},\uv{2}{f}{})+C(\uv{3}{Af}{},\uv{2}{\delta}{})\right]}_{(\lambda_2^f + \alpha_2^\delta)\langle \qdvec, \qv{1}{}{}\rangle} \nonumber\\
    & - \phi\left[\delta \underbrace{\left(C(\uv{3}{\phi}{},\uv{2}{\delta}{})+\Delta \uv{3}{\phi}{}\right)}_{\theta_\delta\langle \qdvec, \qv{1}{}{}\rangle}+f\underbrace{C(\uv{3}{\phi}{},\uv{2}{f}{})}_{\theta_f\langle \qdvec, \qv{1}{}{}\rangle} \right]\\
    & \qquad  \text{ and homogeneous BC on } \partial \Omega, \qquad \qv{5}{}{}\in A_yS_z\nonumber
\end{align}

Note that the condition $\langle \qdvec, \qv{3}{i}{}\rangle=0$ obtained at previous order eliminates the first 3 terms. Using that $d_t a = d_t (\varepsilon A)=  \varepsilon \partial_t A + \varepsilon^3 \partial_T A + \varepsilon^5 \partial_{T_2} A $ we finally obtain the $5^{th}$ order amplitude equation:

\begin{align*}
    d_t a =& -\gamma a^5 - a^3\left(\mu + \underbrace{\varepsilon^2\delta}_{Re_c^{-1}-Re^{-1}} \mu_2^\delta+ \underbrace{\varepsilon^2 f}_{(1+\Gamma_c)^{-1}-(1+\Gamma)^{-1}} \mu_2^f\right) - a^2 \underbrace{\varepsilon^3\phi}_{F} \chi \\
    &- a\left[\varepsilon^2\delta\left( \lambda + \varepsilon^2\delta\lambda_2^\delta  + \varepsilon^2f\lambda_2^f\right) + \varepsilon^2f\left( \alpha + \varepsilon^2\delta\alpha_2^\delta+ \varepsilon^2f \alpha_2^f\right)\right]- \varepsilon^3 \phi \left(\theta +\varepsilon^2\delta \theta_2^\delta+\varepsilon^2f \theta_2^f\right) 
\end{align*}

To plot the fourth panel of fig. \ref{fig:asym} the dependence in $\Gamma$ ($f$) is dropped, as shown in \cref{eq:5thorder}, and only the $3^{rd}$ order effect of asymmetry is kept.

\begin{align}
    d_t a =& -\gamma a^5 - a^3\left(\mu + \varepsilon^2\delta \mu_2^\delta\right) - a\varepsilon^2\delta\left( \lambda + \varepsilon^2\delta\lambda_2^\delta  \right) - \varepsilon^3 \phi\theta. \label{eq:5thorder}
\end{align}

The landau coefficient computed are reported in table \ref{tab: amp_eq} below.

\begin{table}[H]
\begin{tabular}{l|>{\columncolor[HTML]{EFEFEF}}c |>{\columncolor[HTML]{FFFFFF}}c |>{\columncolor[HTML]{EFEFEF}}c |>{\columncolor[HTML]{FFFFFF}}c |>{\columncolor[HTML]{EFEFEF}}c |>{\columncolor[HTML]{FFFFFF}}c |>{\columncolor[HTML]{EFEFEF}}c |}
\cline{2-8} & $\lambda$ & $\mu$                 & $\alpha$ & $\theta$              & $\lambda_2$ & $\mu_2$ & $\gamma$ \\ \hline
\multicolumn{1}{|l|}{\cellcolor[HTML]{FFFFFF}$\Gamma=0$, $Re_c=46$}     & -7.7368   & $8.2262\cdot 10^{-3}$ & 0.608    &  0                   & -           & -       & -        \\ \hline
\multicolumn{1}{|l|}{\cellcolor[HTML]{FFFFFF}$\Gamma=0.1$, $Re_c=72$}   & -4.9047   & $7.09\cdot 10^{-3}$   & 0.452    & $2.997\cdot10^{-2}$ & -           & -       & -        \\ \hline
\multicolumn{1}{|l|}{\cellcolor[HTML]{FFFFFF}$\Gamma=0.15$, $Re_c=95$}  & -3.4625   & $3.00\cdot 10^{-3}$   & .330     & $3.279\cdot10^{-2}$ & -           & -       & -        \\ \hline
\multicolumn{1}{|l|}{\cellcolor[HTML]{FFFFFF}$\Gamma=0.18$, $Re_c=120$} & -2.1762   & $1.42\cdot 10^{-3}$   & .236     & $2.726\cdot10^{-2}$ & -           & -       & -        \\ \hline
\multicolumn{1}{|l|}{\cellcolor[HTML]{FFFFFF}$\Gamma=0.9$, $Re_c=381$}  & -14.367   & $-2.42\cdot 10^{-3}$  & -1.04    & $0.107$            & 12028       & -142.99 & 0.5709  \\ \hline
\multicolumn{1}{|l|}{\cellcolor[HTML]{FFFFFF}$\Gamma=1.1$, $Re_c=204$}  & -44.992   & $9.12\cdot 10^{-2}$   & -1.51    & $2.082\cdot10^{-2}$ & -           & -       & -        \\ \hline
\end{tabular}
\centering
\caption{Landau coefficients up to $5^{th}$ order for different $\Gamma$.}
\label{tab: amp_eq}
\end{table}

\section{Roughness of the channel}\label{annex:roughness}

\begin{figure}[H]
    \centering
    \includegraphics[width=0.5\linewidth]{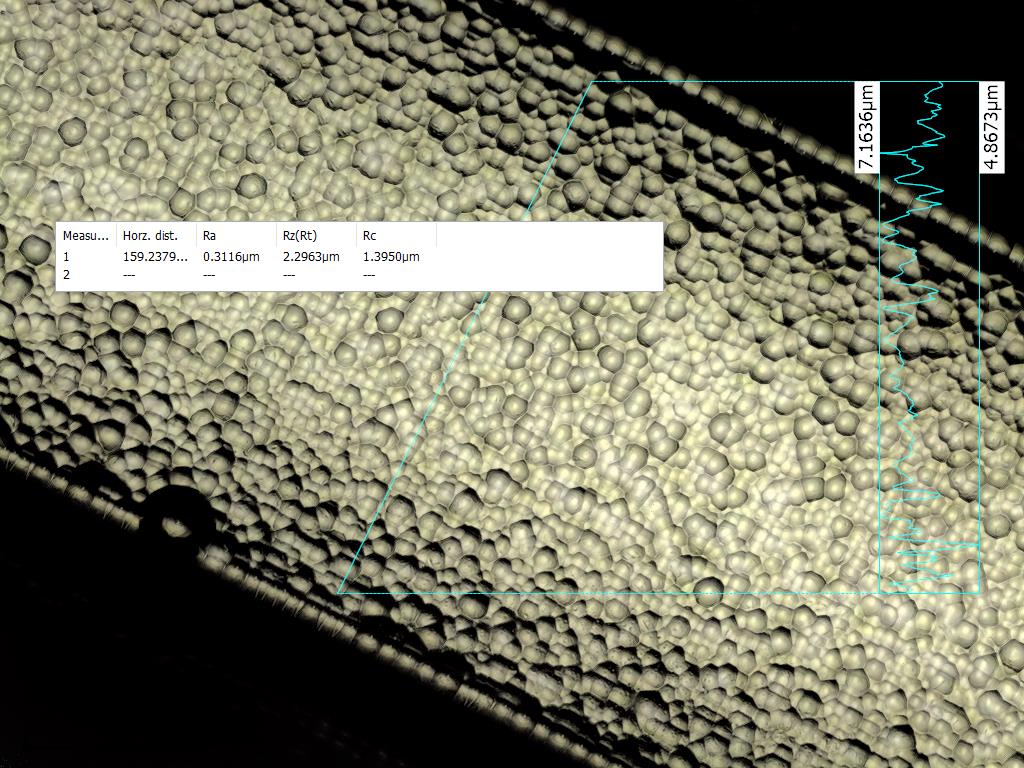}
    \caption{Roughness measurement in the middle of the channel.}
    \label{fig:roughness}
\end{figure}

The roughness has been measured using a Keyence VK-X100 profilometer equipped with a Nikon CF Plan ELWD 50x objective. To determine $R_d$, 3 images (i.e. figure \ref{fig:roughness}) were randomly taken in the bottom of the engraved channel. Their respective roughness was calculated using the Keyence VK-analyzer software and then averaged.
The obtained engraved glass has a root-mean-square roughness deviation of $R_d\simeq 0.8\pm0.1$ $\mu$m, as shown in figure \ref{fig:roughness}.

\section{Inlet profile} \label{annex:inlet}

As explained in the main text, we impose
\begin{equation*}
    \text{at the inlets $\partial\Omega_{c/l_{\pm}}$: }  u(y,z) = C_{c/l_\pm}^{-1} u_{in}(y,z;y_{c/l_\pm},w_{c/l_\pm},H),
\end{equation*}

where $y_{c/l}$ are the center of the inlet faces, $y_{c}=0$, $y_{l_\pm} = \pm \frac{W-w_{l_\pm}}{2}$ and $u_{in}$ is the following dimensionless fully developed profile:

\begin{align}
    &u_{i n}(y, z;y_{c/l_\pm},w_{c/l_\pm},H)=\sum_{\text{n odd}}^{\infty} \dfrac{(-1)^{(n-1)/2}}{n^3}\cos \left(n \pi \frac{y-y_{c/l_\pm}}{w_{c/l_\pm}}\right)\left[1-\frac{\cosh \left(n \pi \frac{z}{w_{c/l_\pm}}\right)}{\cosh \left(n \pi \frac{H}{2 w_{c/l_\pm}}\right)}\right]. \label{annex:uin}
\end{align}

The desired flow rates described in the main text are obtained thanks to the following rescaling coefficients
\begin{align}
    &\begin{cases}
               C^{-1}_c= & hW\dfrac{1}{1+\Gamma}\left[\iint_{\partial\Omega_c} u_{in}(y,z;y_{c/l_\pm},w_{c/l_\pm},H) dydz\right]^{-1},\\\\
               C^{-1}_{l_\pm}=& \dfrac{hW}{2}\dfrac{\Gamma}{1+\Gamma}\left[\iint_{\partial\Omega_{l_\pm}} u_{in}(y,z;y_{c/l_\pm},w_{c/l_\pm},H)  dydz\right]^{-1}.
    \end{cases}\label{annex:C}
\end{align}

\section{$Re_c$ computation, and WNL prediction of $\sigma(Re,\Gamma)$} \label{annexe: LSA WNL}

\begin{figure}[H]
    \centering
    \includegraphics[width=0.8\linewidth]{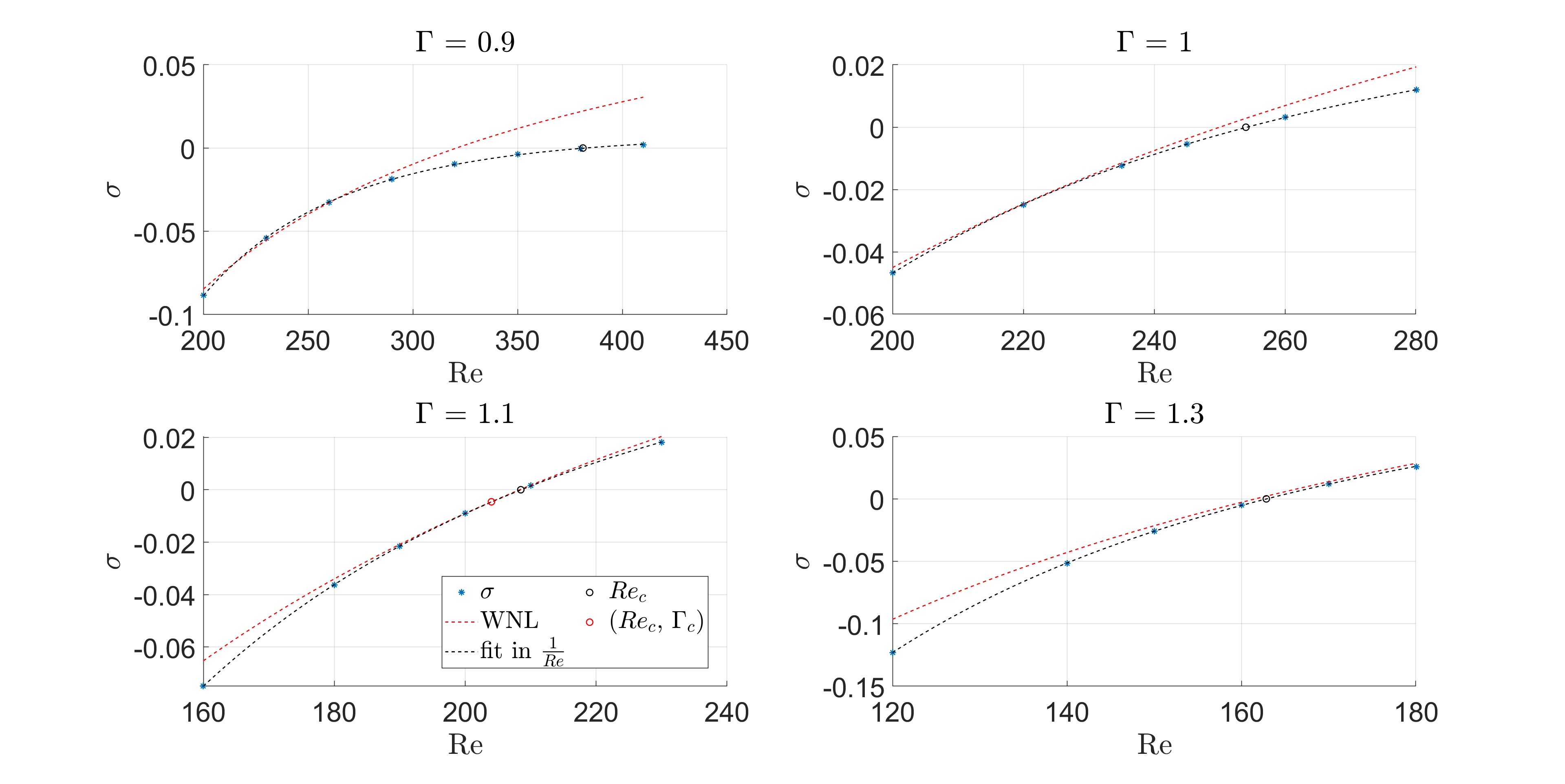}
    \caption{Growth rate $\sigma$ in function of $Re$ in blue asterisks, fit to obtain the critical $Re_c$ in black dashed line and the WNL prediction of $\sigma(Re,\Gamma)$, constructed around ($Re_c=204$, $\Gamma_c=1.1$), the red circle in the bottom left panel, in red dashed lines.}
    \label{fig:pred WNL 1.1}
\end{figure}

To get $Re_c$, instead of performing a large number of computations near the neutral point, we perform a smaller number of computations above and below the neutral point and fit the growth rate (in blue stars) with a quadratic interpolation in $Re^{-1}$ (in black dashed lines in \cref{fig:pred WNL 1.1}). Finding the zero of this polynomial then gives us the value of the critical Reynolds (black circles). Performing the WNL analysis at $(Re_c=204, \Gamma_c=1.1)$, we can get a prediction of the growth rate dependence in $Re$ and $\Gamma$ (red dashed lines). For $\Gamma\in[1,1.3]$ the WNL predicts well the growth rates computed using the LSA which worsens at $\Gamma =0.9$ at higher $Re$. This is because, first of all the WNL prediction is valid for $\varepsilon^2\delta=Re_c^{-1}-Re^{-1}\ll 1$ and $\varepsilon^2 f=(1+\Gamma_c)^{-1}-(1+\Gamma)^{-1}\ll1$ and for $\Gamma =0.9$, $Re=381$, $\varepsilon^2f = -5\cdot10^{-2}$ and $\varepsilon^2\delta = 2\cdot 10^{-3}$ and second of all, at $\Gamma =0.9$ we have a change in the bifurcation nature and, hence, a change in the dependence of the growth rate. 

\end{document}